\documentclass[useAMS,usenatbib,usegraphicx]{mn2e}

\usepackage[T1]{fontenc}
\usepackage{aecompl}
\usepackage{multirow}
\usepackage{amsmath}
\usepackage{hyperref}

\usepackage{color}

\newcommand{\lya}{Ly$\alpha$\ }
\newcommand{\cm}{\, {\rm cm}}
\newcommand{\angs}{ {\rm \AA } }
\newcommand{\kms}{\, {\rm km}\, {\rm s}^{-1} }
\newcommand{\hmpc}{\, h^{-1} {\rm Mpc}}
\newcommand{\msun}{\, {\rm M_\odot} }
\newcommand{\mgii}{Mg\thinspace II }

\title[The Cross-correlation of MgII Absorption and Galaxies in BOSS]{The Cross-correlation of MgII Absorption and Galaxies in BOSS}

\author[Ignasi P\'erez-R\`afols et al]
  {Ignasi ~P\'erez-R\`afols,$^{1,2}$\thanks{email: iprafols@icc.ub.edu}
  Jordi ~Miralda-Escud\'e,$^{1,3}$ Britt ~Lundgren,$^{4,5}$ Jian ~Ge,$^6$
  \newauthor
  Patrick ~Petitjean,$^7$ Donald P.~Schneider,$^{8,9}$ Donald G.~York$^{10}$ 
  \newauthor	
  and Benjamin A.~Weaver$^{11}$\\
  $^1$Institut de Ci\`encies del Cosmos, Universitat de Barcelona/IEEC, Barcelona E-08028, Catalonia\\
  $^2$Departament d'Astronomia i Meteorologia, Facultat de F\'isica, Universitat de Bacelona, E-08028 Barcelona, Catalonia\\
  $^3$Instituci\'o Catalana de Recerca i Estudis Avan\c cats,
Barcelona, Catalonia\\
  $^4$Department of Astronomy, University of Wisconsin - Madison, 475 North Charter Street, Madison, WI 53706 USA\\
  $^5$National Science Foundation Astronomy and Astrophysics Postdoctoral Fellow\\
  $^6$Department of Astronomy, University of Florida, Bryant
Space Science Center, Gainesville, FL 32611-2055, USA\\
  $^7$Institut d'Astrophysique de Paris, UPMC \& CNRS, UMR7095 98bis Boulevard Arago, 75014 - Paris, France\\
  $^8$Department of Astronomy and Astrophysics, The Pennsylvania State University,
   University Park, PA 16802\\
  $^9$Institute for Gravitation and the Cosmos, The Pennsylvania State University,
   University Park, PA 16802\\
  $^{10}$Department of Astronomy and Astrophysics and the
     Enrico Fermi Institute, University of Chicago, \\5640 South Ellis
Avenue, Chicago, IL 60637, USA\\
  $^{11}$Center for Cosmology and Particle Physics, New York University, New York, NY 10003 USA
}

\date{Released 2014 Xxxxx XX}

\pagerange{\pageref{firstpage}--\pageref{lastpage}} \pubyear{2014}

\begin{document}

\maketitle

\label{firstpage}

\begin{abstract}
   We present a measurement of the cross-correlation of MgII absorption and
massive galaxies, using the DR11 main galaxy sample of the Baryon
Oscillation Spectroscopic Survey of SDSS-III (CMASS galaxies), and the
DR7 quasar
spectra of SDSS-II. The cross-correlation is measured by stacking quasar
absorption spectra shifted to the redshift of galaxies that are within a
certain impact parameter bin of the quasar,
after dividing by a quasar continuum model. This results in an average MgII
equivalent width as a function of impact parameter from a galaxy, ranging from
50 kpc to more than 10 Mpc in proper units, which includes all MgII absorbers.
We show that special care needs to be taken to use an unbiased quasar continuum
estimator, to avoid systematic errors in the measurement of the mean stacked
MgII equivalent width. The measured cross-correlation follows the expected
shape of the galaxy correlation function, although measurement errors are
large. We use the cross-correlation amplitude to derive the bias factor of MgII
absorbers, finding $b_{\mathrm{MgII}}=2.33\pm 0.19$, where the error accounts
only for the statistical uncertainty in measuring the mean equivalent width.
This bias factor is larger than that obtained in previous studies and may be
affected by modeling uncertainties that we discuss, but if correct it
suggests that MgII absorbers at redshift $z\simeq 0.5$ are spatially
distributed on large scales similarly to the CMASS galaxies in BOSS.

\end{abstract}


\section{Introduction}  \label{sec: Introduction}

  The \mgii doublet absorption line, at rest-frame wavelengths
$\lambda = 2796.3543$\angs and $2803.5315$\angs\!\!, is an extremely useful
tracer of photoionized gas clouds in galactic halos. Several reasons
account for this trait: magnesium is among the most abundant of the heavy
elements, the oscillator strength of this \mgii doublet is particularly
large, its rest-frame wavelength makes it easily observable from
ground-based telescopes at redshifts $z>0.3$, and magnesium is mostly in
the form of \mgii in photoionized, self-shielded clouds at the typical
pressures of galactic halos. Gas clouds with atomic hydrogen column
densities $N_{HI} \la 2\times 10^{17}\cm^{-2}$ are optically thin
to photons above the hydrogen ionization potential of $13.6$ eV, as well
as to the harder photons that ionize \mgii\!\!. Most of the magnesium is
therefore ionized more than once in thin clouds, greatly reducing the
\mgii column density. On the other hand, in optically thick clouds,
the shielded magnesium ions are mostly recombined to \mgii\!\!, but
MgI has an ionization potential of $7.65$ eV (below that of hydrogen)
and is ionized by photons that penetrate the region of self-shielded
hydrogen. Hence, self-shielded clouds in galactic halos should have
most of their magnesium as \mgii \citep[e.g., ][]{Bergeron&Staniska1986}.
In fact, most strong \mgii systems
are also Lyman limit systems, i.e., their HI column densities are high
enough to be self-shielding \citep{Rao2006}.

  The rest-frame equivalent width distribution of \mgii absorption
systems is approximated by an exponential form,
${\rm d}N/{\rm d}z \propto \exp(-W/W^{*})$, with a value
$W^{*} \simeq 0.6 \angs $ at $z=0.5$ that increases
gradually with redshift, and an excess of systems over this form at
$W< 0.3$\angs \citep[and references therein]{Nestor2005,Narayanan2007}. 
Most of the strong systems have a complex velocity 
structure, with a velocity dispersion of 
multiple absorbing components that is characteristic of galaxy velocity
dispersions, favoring models of a collection of photoionized clouds
randomly moving through a galactic halo 
\citep{Bahcall1975,Sargent1979,Churchill2000}, as had already been proposed by \cite{Bahcall&Spitzer1969}.

  The association of \mgii absorption systems with galactic halos was
firmly established with the work of \cite{Lanzetta&Bowen1990},
\cite{Bergeron&Boisse1991}, and \cite{Steidel1994}.
The observations of the frequency of occurrence of \mgii absorbers at
different impact
parameters from luminous galaxies led to a simple model of halos that
are close to spherical, in which absorption with rest-frame equivalent
width $W > 0.3$\angs is nearly always observed within an impact
parameter $r_{\rm p} \la 50 (L_{\rm K}/L_{\rm K}^{*})^{0.15}\, {\rm kpc}$
of a galaxy of K-band luminosity $L_K$, and becomes rapidly weaker at
larger radii, independently
of the type of galaxy being considered \citep{Steidel1995}. Actually,
the natural expectation is that there is a smooth profile of declining
mean \mgii absorption strength with impact parameter around a galaxy,
caused by a decreasing density of clouds with radius in a galactic halo.
This is consistent with more recent work, where the
mean \mgii equivalent width (which is indicative of the number of
intersected individual clumps with saturated absorption) has been shown
to steeply decline with the impact parameter $r_p$ roughly as
$\overline{W} \propto r_{\rm p}^{-1.5}$ \citep{Chen2010a},
with a characteristic radius at a fixed $\overline{W}$ that scales
proportionally to $R_{\rm \mgii}\propto {M_{*}}^{0.3} (\mathrm{sSFR})^{0.1}$
\citep{Chen2010b}, where $M_{*}$ is the stellar mass in the galaxy 
and sSFR is the star formation rate per unit of stellar mass. We also
note that this describes the mean profile of MgII absorption around
galaxies, and that there are rare cases where MgII absorption is absent
even at very small impact parameter from luminous galaxies
\citep{Johnson2014}.

  The large extent of the gaseous halos traced by metal absorbers and
their nearly ubiquitous presence around all massive galaxies, with only
a weak dependence on the specific star formation rate, are observational
facts providing support to a scenario in which the \mgii absorbers
are a signature of the accretion process of new material onto galaxies.
Accreting gas at the temperatures of virialization in galactic halos is
thermally unstable and should naturally form photoionized clouds
whenever the cooling time of hot halo gas is short compared to the age
of the galaxy. This behaviour naturally leads to a two-phase model of gaseous
galactic halos, where cool clouds can form in approximate pressure
equilibrium with a hot medium and are produced in abundance within the
cooling radius
\citep{Spitzer1956,Mo&Miralda-Escude1996,Maller&Bullock2004}.
On the other hand, there is evidence at small radii
pointing to the impact of galactic winds on \mgii absorbers:
the absorbers are more numerous near the minor axis of their associated
galaxies \citep{Bordoloi2011,Bordoloi2012, 
 Kacprzak2011,Kacprzak2012,Lundgren2012}.
The distribution of \mgii absorption
systems around galaxies is therefore likely to be sensitive to
processes involving both inflow and outflow of material into and from
the regions containing the bulk of the stellar mass in galaxies.

  The association of \mgii absorbers with galaxies implies a large-scale
cross-correlation of these objects. The cross-correlation is in general
a result of two effects. First, if every galaxy is surrounded by a gas
halo, a \mgii absorber located at an impact parameter $r_{\rm p}$ from a
galaxy may actually be associated with this galaxy, and be part of the
gas halo around it. Second, the \mgii absorber may be associated with a
different galaxy that may be a satellite of the first, or with an
unrelated galaxy that is spatially correlated with the first, following
the usual galaxy autocorrelation. These are usually described as the
1-halo and 2-halo terms, although MgII systems are considered to be
associated to galaxies, rather than dark matter halos. In the limit of
impact parameters much smaller than the typical size of a galaxy halo,
the first term, determined by the gas halo profile around each galaxy
\citep[e.g.,][]{Tinker&Chen2008}, dominates, whereas in the limit of
large impact parameters, the second term is the important one. The total
cross-correlation is in general a combination of the two terms over a
wide range of intermediate impact parameters, and it is impossible to
cleanly separate the two contributions. But in the large-scale limit,
the cross-correlation of \mgii absorbers and galaxies should follow the
form of the galaxy correlation function, with an amplitude that is
proportional to the product of the two bias factors of the two
populations. Hence, measuring the large-scale clustering amplitude helps
determine the bias factor, and therefore the halo population that the
\mgii absorbing clouds are associated with.

  The large-scale \mgii\!\!-galaxy cross-correlation was first measured
using the photometric catalog of Luminous Red Galaxies in the Sloan
Digital Sky Survey \citep[hereafter, SDSS; see][]{York2000} and the set of individually
detected \mgii absorbers in the spectra of SDSS quasars by 
\cite{Bouche2004}, \cite{Bouche2006}, \cite{Lundgren2009}, \cite{Gauthier2009} and \cite{Lundgren2011}.
In the absence of precise galaxy
redshifts, only the projected cross-correlation function can be
measured. 
The work by \cite{Lundgren2009}, based on a set of 2705 \mgii 
absorbers with rest-frame equivalent width $W> 0.8$\angs
over the redshift interval $0.36 < z < 0.8$, found that the form of the
projected cross-correlation is well matched by the Luminous Red Galaxy
auto-correlation in the impact parameter range
$0.3 \hmpc < r_p < 30 \hmpc$, and measured
the bias factor of these \mgii systems to be $b_{\mathrm{Mg}}=1.10 \pm 0.24$.
\cite{Gauthier2009} performed a similar analysis for a sample of 
1158 \mgii absorbers with $W <1$\angs over the redshift interval 
$0.4 < z < 0.7$ and found the bias factor of these \mgii systems 
to be $b_{\mathrm{Mg}}=1.36 \pm 0.38$.
Both works found an indication that weak absorbers, with $W < 1.5$\angs,
cluster more strongly (i.e., they have a larger bias factor) than
strong absorbers, and are therefore located in more massive halos on
average, although this result was not of high statistical
significance. The method used by \cite{Bouche2006}, \cite{Lundgren2009} and \cite{Gauthier2009},
 based on a catalog of identified \mgii systems, requires
careful attention to the selection function of {\it both}
galaxies and \mgii systems through an extensive use of simulations, because
the number of absorbers will obviously be enhanced in regions of the
survey containing more galaxies and more quasar spectra, or where the
spectra are of higher signal-to-noise ratio, owing to variable observing
conditions and also intrinsic clustering of the quasar sources.

  We present a different approach in this paper to measure the
cross-correlation of massive galaxies and \mgii absorbers, based on
the galaxies with spectroscopically-measured redshifts of the new
Baryon Oscillation Spectroscopic Survey (hereafter, BOSS) of the 
SDSS-III Collaboration \citep{Eisenstein2011,Dawson2013}.
Instead of identifying
individual \mgii absorbers, we use a stacking method to measure the
average \mgii absorption around a galaxy as a function of the impact
parameter and redshift separation. In other words, we measure the
redshift-space cross-correlation function of galaxies and \mgii
absorption. Our approach is similar to that of \cite{Zhu&Menard2013b},
who have measured the mean CaII absorption around galaxies,
and to \cite{Zhu2013}, who have obtained a similar
measurement of the large-scale \mgii absorption; we compare their results
with ours and discuss the differences near the end of this paper.
The data set we use is described in Section \ref{sec: Sample data} , and the
method is presented in detail in Sections \ref{sec: Stacking procedure} and 
\ref{sec: absorption model}. We present the results in Section \ref{sec: Results}, 
which are applied to infer the mean bias factor of \mgii systems in Section 
\ref{sec: Discussion}. Finally we summarize our conclusions in Section 
\ref{sec: Conclusions}. Throughout this paper we use the $\Lambda$CDM model 
with $H_{0}=68 \kms \,{\rm Mpc}^{-1}$ and $\Omega_{m}=0.3$.

\section{Data Sample}  \label{sec: Sample data}

  The first step in the analysis is to identify quasar-galaxy pairs in which
the quasar sightline passes within a specified bin of projected proper radius,
or impact parameter $r_{\rm p}$, from the foreground galaxy.
For the background quasar sample we use the quasar catalogue of \cite{Schneider2010} from the 7th Data Release \citep[DR7,][]{Abazajian2009} of the SDSS-II Collaboration \citep{Gunn1998,York2000,Gunn2006,Eisenstein2011,Bolton2012,Smee2013}, with 105,783 spectroscopically confirmed quasars. 
For the galaxies, the CMASS catalogue \citep{Dawson2013} of the SDSS-III Collaboration that is prepared for the 11th Data Release (DR11, extension of the SDSS DR9 
\citep{Ahn2012} and the SDSS DR10 \citep{Ahn2013}) is used, which contains a
total of 938,280 galaxies. The DR11 galaxies represent the majority of the final
BOSS sample, also covering most of the sky area included in DR7.
We note that this galaxy catalogue is the same that was used by
\cite{Zhu2013}, even though they refer to it as the Luminous Red Galaxy
catalogue.
We exclude any galaxies at redshift lower than $0.35$, corresponding to
$1+z=\lambda_{\rm min}/\lambda_{\mathrm{MgII}}$, where we set
$\lambda_{\rm min}=3800$\angs as the shortest wavelength with sufficient
signal-to-noise ratio to provide a useful \mgii absorption signal.
This requirement reduces the galaxy sample to 895,472 galaxies.
The quasar sample could be increased by including the DR11 sample from
BOSS, but most of the new quasars in DR11 are fainter than those in DR7
(having therefore lower S/N) and are at a high redshift, at which the
\mgii lines associated with the CMASS galaxies appear superposed with the
\lya forest.

The mean \mgii absorption around galaxies is computed by stacking all
the spectra within a certain range of impact parameters from a galaxy.
In general, all quasars probing a line-of-sight within a maximum proper
impact parameter that was set to $r_{\rm p,max}=12.8$ Mpc are included in the
spectral stacking, provided that the following
restrictions are met: first, in order to avoid broad absorption line
systems associated with quasar outflows, the redshift of the quasar,
$z_{\rm q}$, must be larger than that of the galaxy, $z_{\rm gal}$, by a
minimum amount, corresponding to a velocity $v=0.04c$,
\begin{equation}
	\label{eq: redshift restriction 1}
	1+z_{\rm q} > R (1+z_{\rm gal}) ~,
\end{equation}
where $R=\sqrt{(1+v/c)/(1-v/c)}\simeq 1.041$. Second, because the
presence of
the \lya forest substantially increases the noise in quasar spectra,
pairs where the galaxy \mgii line would fall in the \lya forest region
of the quasar are also excluded; in other words, we require
\begin{equation}
	\label{eq: redshift restriction 2}
	(1+z_{\rm gal})\lambda_{\mathrm{MgII}} > (1+z_{\rm q})\lambda_{\alpha} ~,
\end{equation}
where $\lambda_{\mathrm{MgII}} = 2798.743$\angs is the mean
wavelength of the \mgii doublet, and the hydrogen Ly$\alpha$ wavelength
is $\lambda_{\alpha} = 1215.67$\angs at the rest-frame.
Note that the quasars that have been reported to have broad absorption
lines are also
included in the sample. We tested the effect of removing them and we
found that the measurements are not significantly changed. We believe
that this is because we are only using small windows on the quasar
spectra, and therefore the broad absorption line systems are typically
not included in the analysis.

\begin{figure*}
	\includegraphics[width=\textwidth]{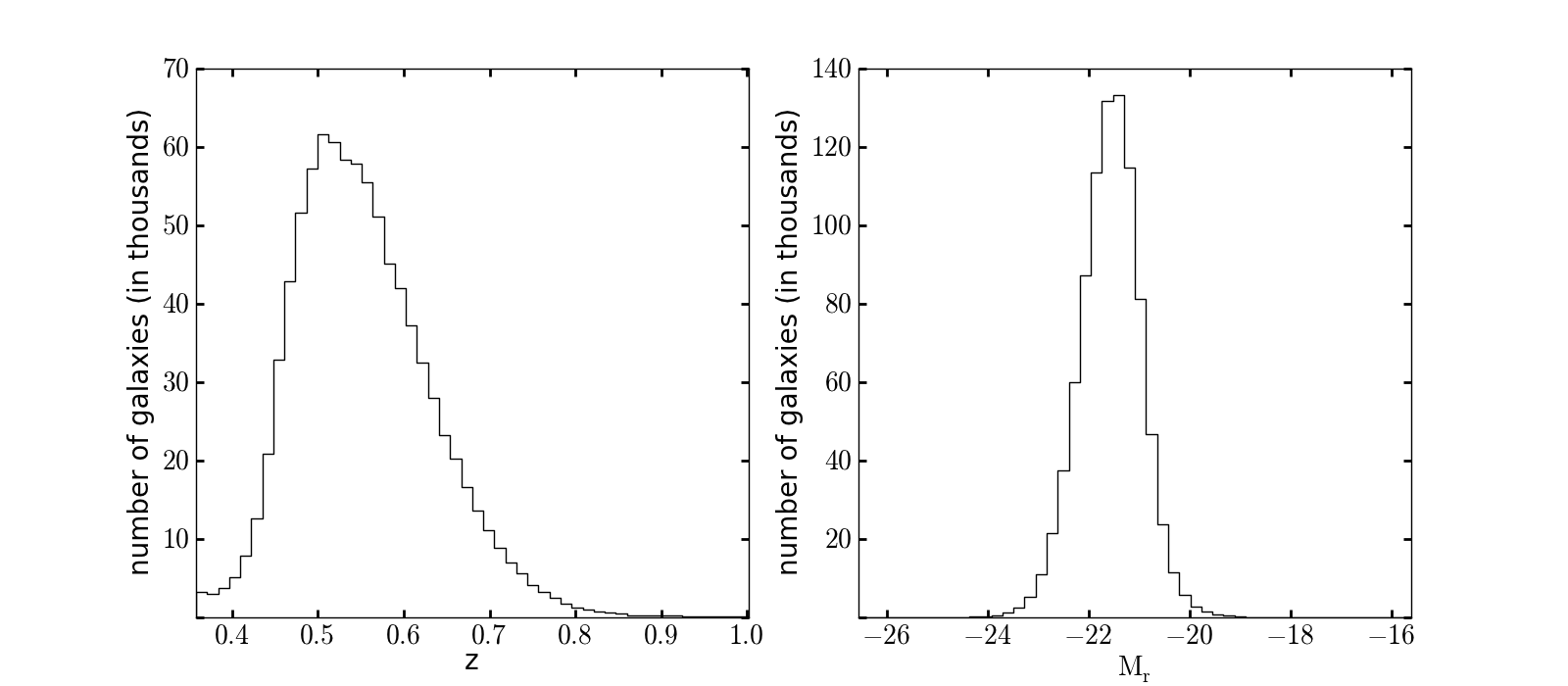}
	\caption{Redshift (left pannel) and luminosity (right pannel)
distributions for the selected CMASS galaxies (see text for details) .}
	\label{fig: histograms}
\end{figure*}

  The luminosity and redshift distributions of all the DR11 CMASS
galaxies meeting these two conditions for at least one quasar that is
within an impact parameter $b_{\rm max}=12.8$ Mpc, as computed from the angular
separation at the redshift of the galaxy, are shown in figure
\ref{fig: histograms}. Note that the density of quasars
in DR7 implies that each galaxy has on average $\sim 3$ quasars within this
impact parameter, therefore the majority of galaxies have at least one
quasar pair and these distributions are nearly the same as those of the
whole DR11 CMASS sample. These distributions represent the
characteristics of the galaxies for which we measure the mean \mgii
absorption equivalent width as a function of impact parameter in this
paper.

\section{Stacking procedure} \label{sec: Stacking procedure}

  This section describes the method used to measure the average \mgii
absorption equivalent width as a function of impact parameter from the
CMASS galaxies. In general, the mean transmission fraction $\overline{F}$ of
light from a background quasar due to \mgii absorption line systems
can be written as a function of the impact parameter $r_{p}$ and velocity
separation $v$ from a galaxy, as
\begin{equation}
 \label{eq: deltamg}
 \overline{F}(r_{\rm p},v)= \exp\left[-\tau_{\rm e}(r_{\rm p},v)\right]=
 \exp\left[ -\tau_{\rm e0}(1+\delta_{\mathrm{Mg}}(r_{\rm p},v)) \right] ~.
\end{equation}
Here, $\tau_{\rm e}$ is the effective optical depth, and $\tau_{\rm e0}$ is its
average value at any random position, irrespective of the presence of a
galaxy. The perturbation $\delta_{\mathrm{Mg}}$ is the relative increase of the
effective optical depth of \mgii absorption associated with the presence
of a galaxy at impact parameter $r_{\rm p}$ and velocity separation $v$. The
shape of this perturbation as a function of $v$, for a given
impact parameter, is rather complicated because it arises from the
distribution of relative velocities of a set of doublet lines with
different degrees of saturation; besides, the observations are
affected by the instrumental resolution and the binning. We will return
to these details later.

  The mean effective optical depth at a random position can be expressed in
terms of the rest-frame equivalent width distribution of \mgii absorbers, as
\begin{equation}
	\label{eq: tau_e0}
 \tau_{\rm e0}=\int_{0}^{\infty} {\rm d}W \, \frac{\partial^{2}
 \mathcal{N}}{\partial W\partial z}\, (1+z) \frac{W[1+\bar q(W)]}
{\lambda_{\rm{\mgii}}} ~,
\end{equation}
where the equivalent width $W$ is that of
the strongest line of the \mgii doublet at
$\lambda=2796.35$\angs\!\!, and
$\bar q(W)$ is the mean equivalent width ratio of the two doublet lines.

  Our aim in this paper is to measure the excess in the effective
optical depth,
\begin{equation}
	\label{eq: deltatau}
 \delta\tau_{\rm e}(r_{\rm p},v) = \tau_{\rm e}(r_{\rm p},v)-\tau_{\rm e0} =
 \tau_{\rm e0} \delta_{\mathrm{Mg}}(r_{\rm p},v) ~,
\end{equation}
which is induced by the presence of a galaxy at impact parameter $r_p$ and
velocity separation $v$. This quantity is directly related to the
cross-correlation of \mgii clouds with CMASS galaxies.
We will focus in particular on the projected
value of $\delta\tau_{\rm e}$, obtained after integration over velocity, and its relationship
to the projected cross-correlation. The method we use to measure
this cross-correlation is to average the transmitted fraction over a
large number of lines-of-sight within a given range of impact parameter,
in order to reduce the photon noise and the noise arising from quasar
continuum variability.

  A crucial step to measure the \mgii\!\!\!-galaxy cross-correlation, in the
form of $\delta\tau_{\rm e}$, is to estimate the quasar continuum with a
method that is, to the best possible degree, free of systematic errors
when averaging over a large number of lines-of-sight. In particular,
it is important to ensure that
the presence of a \mgii line itself, which in most cases may be too weak
to be individually detected in the relatively low signal-to-noise ratio SDSS
quasar spectra, does not bias the estimate of the continuum. Obviously,
if the spectral region where we expect to find the \mgii line associated
with a galaxy is used for the continuum determination, the
continuum estimate may be systematically biased too low because of the
presence of an undetected \mgii line. This systematic bias may be
important when stacking large numbers of spectra, even though the \mgii
lines causing the bias are not individually detected in any single
spectrum.

  To illustrate the importance of the quasar continuum
determination for the problem of measuring the average \mgii absorption
around galaxies at large impact parameters, we explore two different methods
and we perform a number of tests for the presence of systematic errors.
The first method, designated as {\it mean subtraction}, is specifically
designed for our problem, while the second method, designated as
{\it variable smoothing},
fits a continuum by iteratively smoothing the spectrum with a variable
smoothing length to flexibly fit both emission lines and featureless
continuum regions. The latter method uses the entire spectrum to
determine the continuum, including the region where the associated
\mgii line is expected, and is therefore subject to the
systematic error described above. We now describe each method in detail.
Tests of the methods that show that the mean subtraction method
correctly recovers the mean equivalent width and the variable
smoothing method is subject to various systematic errors, including the
continuum fitting bias mentioned above, are presented in Appendix \ref{sec: Test}.

\subsection{Method 1: Mean subtraction} \label{subs: Mean subtraction}

  The first approach uses the mean spectrum of all quasars as a
continuum fit model. Each quasar spectrum is divided by the mean quasar
spectrum, and then a linear fit to this ratio is obtained
around the spectral region of the expected \mgii line 
for each galaxy-quasar pair,
but without using the narrower central interval where the \mgii line should
be located. This linear fit is used to further improve the continuum
estimate, allowing for intrinsic variations of the quasar continua. The
results are then stacked for all quasar-galaxy pairs at each interval of
impact parameter, after shifting to the redshift of the galaxy in each
pair, to obtain the final composite \mgii absorption spectrum. We now
explain each of these steps in detail.

\subsubsection{Generating the mean spectrum}
 \label{subsubs: Generating mean spectrum}
 
  The mean quasar spectrum is generated using the DR7 quasar spectra
following a similar approach as the one undertaken by
\cite{VandenBerk2001}. The mean spectrum in \cite{VandenBerk2001} was
generated using 1,800 quasars \citep[figure 3 of][]{VandenBerk2001}, so our
mean spectrum is more accurate because of the much larger number of
quasars available in DR7. In addition, we normalize the spectra using a
spectral window that is particularly suited to obtaining the most
accurate continuum  model in the region where \mgii absorption lines are
found. There is therefore a small difference between our mean spectrum
and that of \cite{VandenBerk2001}. 

\begin{figure*}
	\includegraphics[width=\textwidth]{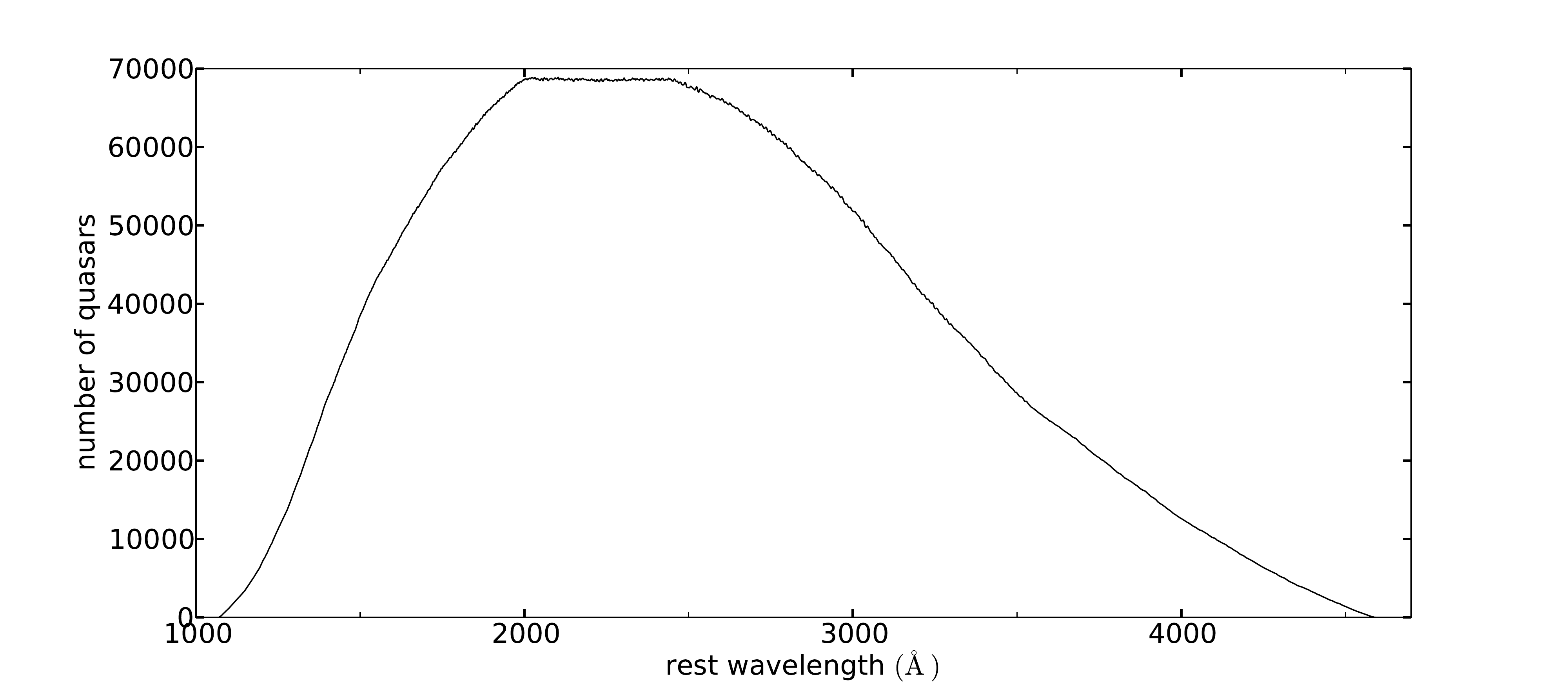}
	\caption{Number of quasar spectra contributing to each
$1$\angs bin of the mean spectrum, as a function of rest-frame
wavelength. The flat feature corresponds to the
spectral range used to compute the normalizing factor $n_j$. Outside
this range the number of quasars contributing to the mean quasar
spectrum decreases because some of the quasar spectra do not extend to
that wavelength.
}
	\label{fig: QSO_mean_counts}
\end{figure*}

  The quasar spectra are first shifted to the rest-frame, using the
redshifts of \cite{Schneider2010}, and rebinned into a common wavelength
scale of $1$\angs per bin, which is close to the resolution
of the observed spectra when shifted to the rest-frame. The values of
the flux and the error at each pixel in the new binning are determined
by the average values of the flux and error in the original pixels that
are projected, partly or fully, to the new pixel, weighted by the
fraction of the new bin covered by each original bin. Each quasar
spectrum, denoted by an index $j$, is normalized with a normalizing
factor $n_{j}$ equal to the mean flux in the interval $2000-2600$\angs\!\!,
\begin{equation}
	\label{eq: norm_factor}
	n_j = \sum_{i}f_{ij}/N_j ~ .
\end{equation}
where $f_{ij}$ is the measured flux value at pixel $i$ of spectrum $j$,
and $N_j$ is the number of pixels in the rest-frame wavelength interval
$2000$\angs$ < \lambda_{ij} (1+z_j) < 2600$\angs\!\!.
Any quasar spectra that do not cover this entire range
of rest-frame wavelength are discarded. The final number of quasar
spectra that are averaged in each $1$\angs bin is shown in
figure \ref{fig: QSO_mean_counts}. The flat top corresponds to the
spectral window used to compute the normalizing factor $n_j$. Note that
the total number of quasar spectra used in this method is $70,650$.
The maximum number of quasars shown in figure
\ref{fig: QSO_mean_counts}, roughly $68,600$ quasars, is lower than the
total number of spectra used because we remove pixels affected by sky
lines. Even though these sky lines are corrected by the BOSS pipeline,
the noise in the affected pixels may sometimes not be well characterized,
so it is best to remove these pixels. For each sky line,
we remove a set of neighboring pixels following the algorithm summarized
in \cite{Palanque-Delabrouille2013}
\citep[for a more detailed explanation of the algorithm
refer to][]{Lee2013}. 

\begin{figure*}
	\centering
	\includegraphics[width=\textwidth]{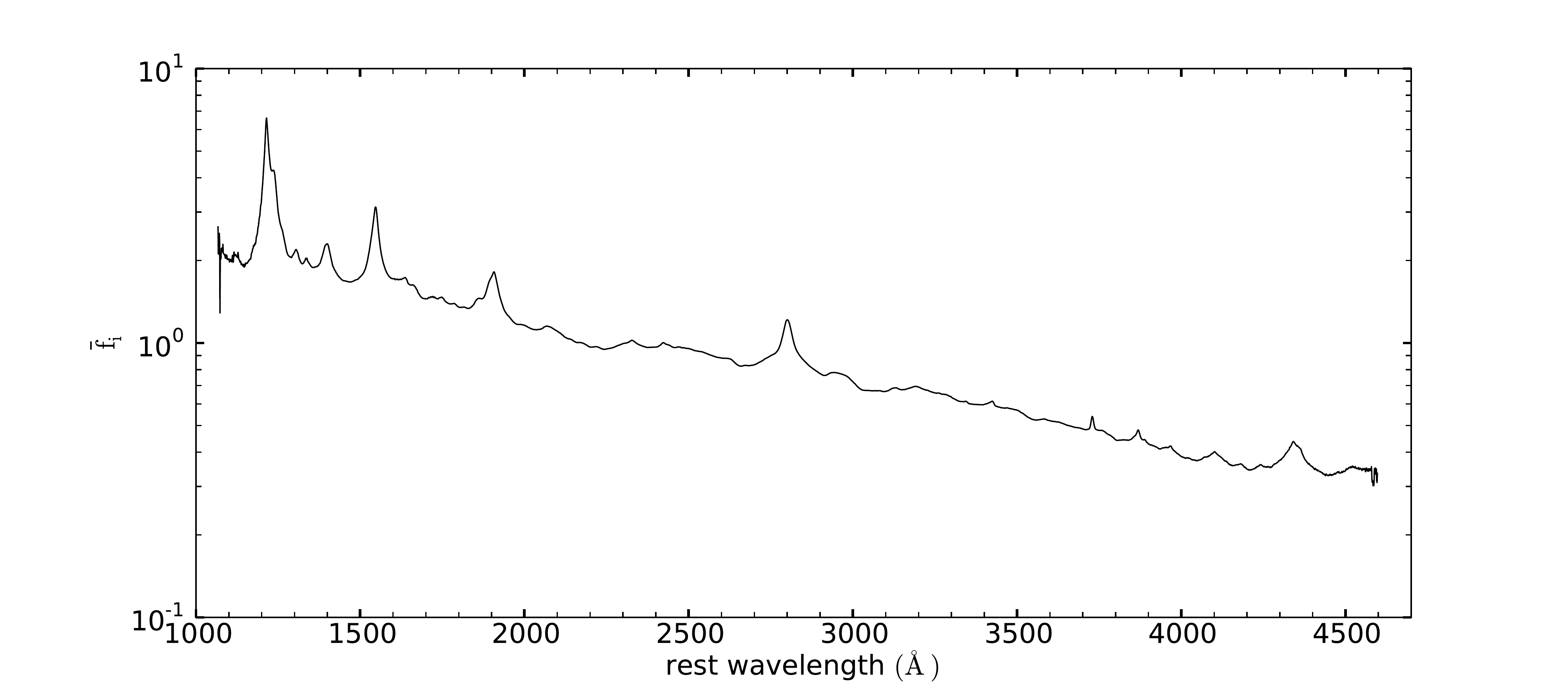}
	\caption{Mean spectrum of the weighted-average obtained from the $70,650$
DR7 quasars, normalized in the rest-frame wavelength interval from $2000$
to $2600$\angs. This mean spectrum is used as a continuum
model.}
	\label{fig: QSO_mean_spectrum}
\end{figure*}

  The rest-frame wavelength interval of $2000 - 2600$\angs is
also used to assign a mean signal-to-noise ratio $s_{j}$ to each
spectrum, calculated as
\begin{equation}
	\label{eq: SNR}
	s_j = \frac{\sum_{i}f_{ij}/N_j}
{\left(\sum_{i}e_{ij}^{2}/N_j\right)^{1/2}} ~ ,
\end{equation}
where $e_{ij}$ is the noise of the flux $f_{ij}$. The mean,
normalized quasar spectrum is then
obtained as a weighted average of all the quasar spectra,
\begin{equation} 
	\label{eq: mean flux}
 \overline{f}_i = \frac{\sum_j\,w_j\left(f_{ij}/n_j\right)}{\sum_j\,w_j} ~,
\end{equation}
where the weights $w_j$ are set equal to
\begin{equation} 
	\label{eq: weight}
	w_j = \frac{1}{s_j^{-2}+\sigma_{I}^{2}} ~.
\end{equation}
The constant $\sigma_I$ is introduced to avoid the highest
signal-to-noise ratio spectra from excessively contributing to the final
average, taking into account the presence of intrinsic quasar spectral
variability, while reasonably weighting down the more noisy spectra. We
fix this constant to $\sigma_I=0.05$ (a reasonable estimate for the
typical fractional intrinsic variability) throughout this paper. 
The resulting mean spectrum is shown in figure \ref{fig: QSO_mean_spectrum}.

\subsubsection{Generating the composite \mgii absorption spectra}
 \label{subsubs: Generating composite spectra} 

  The composite \mgii absorption spectra are obtained by stacking the
spectra of all the quasars having a galaxy within the corresponding impact
parameter bin, after being shifted to the rest-frame of the galaxy in
the region around the \mgii doublet wavelength. For each quasar-galaxy
pair, the quasar spectrum is first rebinned in a velocity variable $v$,
defined in terms of the wavelength separation from the \mgii absorption
line in the rest-frame of the galaxy at redshift $z_{\rm gal}$,
\begin{equation}
	\label{eq: velocity transform}
	v = c\cdot\frac{\lambda-\lambda_{0}}{\lambda_{0}} ~,
\end{equation}
where $\lambda_{0}=\lambda_{\rm \mgii}(1+z_{\rm gal})$, and
$\lambda_{\rm \mgii} = 2\,798.743\, {\rm \AA}$. We use the same linear
rebinning method described in Section
\ref{subsubs: Generating mean spectrum}, with a bin size
$\Delta v = 50\, {\rm \kms}$. The mean spectrum is also rebinned in the
same manner, but using
$\lambda_{0}=\lambda_{\rm \mgii} (1+z_{\rm gal})/(1+z_{\rm q})$, where $z_{\rm q}$ is the
quasar redshift, because the mean quasar spectrum is computed in the
quasar rest-frame.

  The rebinned spectra $f^{(r)}_{ik}$, where the $i$ index now labels
the new bins in $v$, and the $k$ index labels each quasar-galaxy pair in
a certain impact parameter bin, are then divided by the continuum to
obtain a first estimate of the transmission fraction $F^{(0)}_{ik}$,
\begin{equation}
	\label{eq: normalization 1 flux}
	F^{(0)}_{ik}=\frac{f^{(r)}_{ik}/n_{j(k)}}{\overline{f}_{i}} ~,
\end{equation}
where the normalization factor $n_{j(k)}$ is that of the $j$ quasar
corresponding to each pair $k$. The mean quasar spectrum $\overline{f}$
is understood to be the rebinned one and evaluated at the same bins in
$v$ for each quasar-galaxy pair. Hence, if all quasars had identical
intrinsic emission spectra, and in the absence of intervening absorption
and observational noise, this transmission would be equal to unity for
all quasars. The errors are normalized in the same way and computed
according to $E_{ij} = e_{ij}/n_{j}/\overline{f}_{i}$.

  In order to account for intrinsic variations in the spectra of
quasars, we allow for a local smooth gradient in the ratio of each
quasar spectrum to the mean spectrum in the region around each expected
\mgii absorption line. This is modelled by first calculating a weighted average
value of $F^{(0)}_{ik}$ on two intervals in $v$ on each side of the expected
\mgii absorption associated with the galaxy, which are far enough from
the center so that any associated absorption can be neglected. The
intervals used throughout this paper are
$-5\,000 \kms < v < -2\,000 \kms$ and $2\,000 \kms < v < 5\,000\kms$,
with their weighted averages designated as $F^{(-)}_k$ and $F^{(+)}_k$,
respectively. The weights for each bin are set to
$w_{ik}=(\sigma_{n,ik}^2+\sigma_I^2)^{-1}$, where
$\sigma_{n,ik}=E_{ik}/F^{(+,-)}_k$ is the inverse signal-to-noise ratio at
each pixel (we use the averages $F^{(+)}_k$ and $F^{(-)}_k$ for the
signal, instead of the values at each pixel $F^{(0)}_{ik}$, to avoid
biasing the result by systematically giving higher weights to pixels
with positive noise).
We use, as before, $\sigma_I=0.05$.
Weighted averages of the mean velocities $v^{(-)}_k$ and
$v^{(+)}_k$ are computed in the same manner, which are usually close to the central
values of the intervals ($-3500$ and $+3500 \kms$), but not exactly so.
A linear function $L_{ik}$ matching these two points is then defined,
\begin{equation}
	\label{eq: linear}
        L_{ik}= F^{(-)}_k + (F^{(+)}_k - F^{(-)}_k)(v_i-v^{(-)}_k) / (v^{(+)}_k-(v^{(-)}_k) ~.
\end{equation}
In order to better adjust the quasar continuum in the presence of
unrelated random absorption lines or bad pixels, the calculation of the
two weighted averages $F^{(-)}_k$ and $F^{(+)}_k$ is recomputed after
eliminating all outlier pixels in which the normalized flux deviates by
more than 3-sigma from the fitted linear function, i.e., pixels where
$\| F^{(0)}_{ik}-L_{ik} \| > 3(\sigma_{n,ik}^{2}+\sigma_{I}^{2})^{1/2}$. If
more than $20\%$ of the pixels in any of the two intervals are rejected
under this criterion, the quasar spectrum is considered anomalous in the
region of the expected \mgii line and the quasar-galaxy pair under
consideration is rejected and not included in the final processing.

  The transmission fraction is then corrected by this linear fit as
\begin{equation}
	\label{eq: normalization 2}
	F_{ik} = F^{(0)}_{ik}+\left(1-L_{ik}\right) ~.
\end{equation}
We note here that although it would be in principle more correct to
divide by the linear fit, setting
$F_{ik} = F^{(0)}_{ik}/L_{ik}$, we found that this procedure inevitably
introduces a systematic feature in the final stacked spectrum owing
to the fact that a Gaussian error in the function $L$ results in a
non-gaussian distribution of the final transmission $F$ when $L$ is in
the denominator, which is very difficult to correct for. We therefore
decided to subtract $L$ following equation (\ref{eq: normalization 2}).

  Finally, we use the same weights as in equation (\ref{eq: weight}) to
compute the final composite spectrum and its errors,
\begin{equation}
	\label{eq: mean flux 2}
  \overline{F}_i = \frac{\sum_{k}F_{ik}w_{j(k)}}{\sum_{k}w_{j(k)}} ~;
 \qquad
  \overline{E}_{i}^{-2} = \frac{\sum_{k}E_{ik}^{-2}w_{j(k)}}
   {\sum_{j}w_{j(k)}} ~.
\end{equation}
The index $j(k)$ refers to the quasar index $j$ corresponding to each
quasar-galaxy pair labeled by the index $k$.
In these sums, any pixels over the interval from $-5000$ to $+5000\kms$
with a normalized flux $F_{ik}$ below $-2$ or above $+3$ are eliminated, to
exclude bad pixels or values that may have excessive noise.
This eliminates only 0.02\% of the pixels.

\begin{figure*}
	\includegraphics[width=0.5\textwidth]{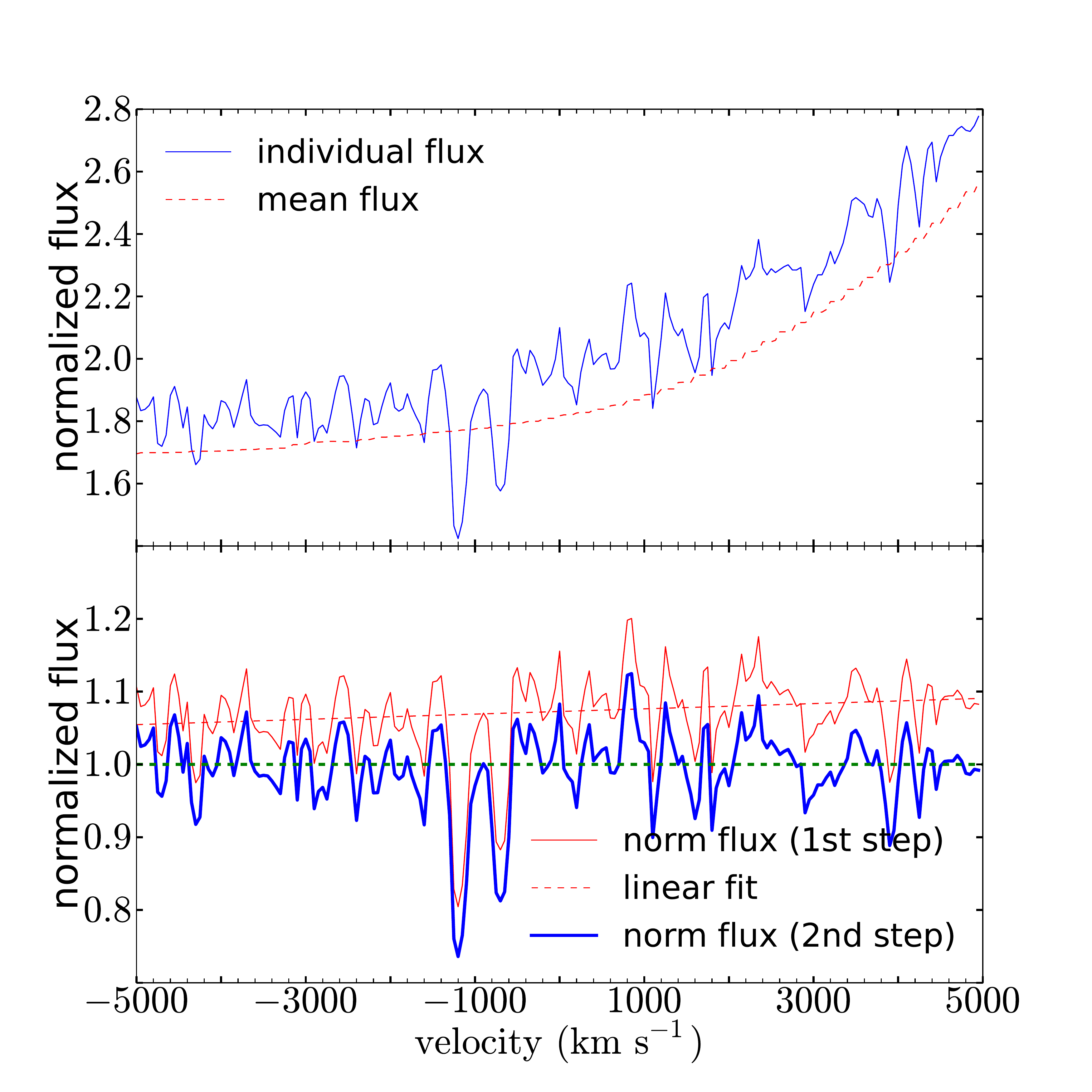}\includegraphics[width=0.5\textwidth]{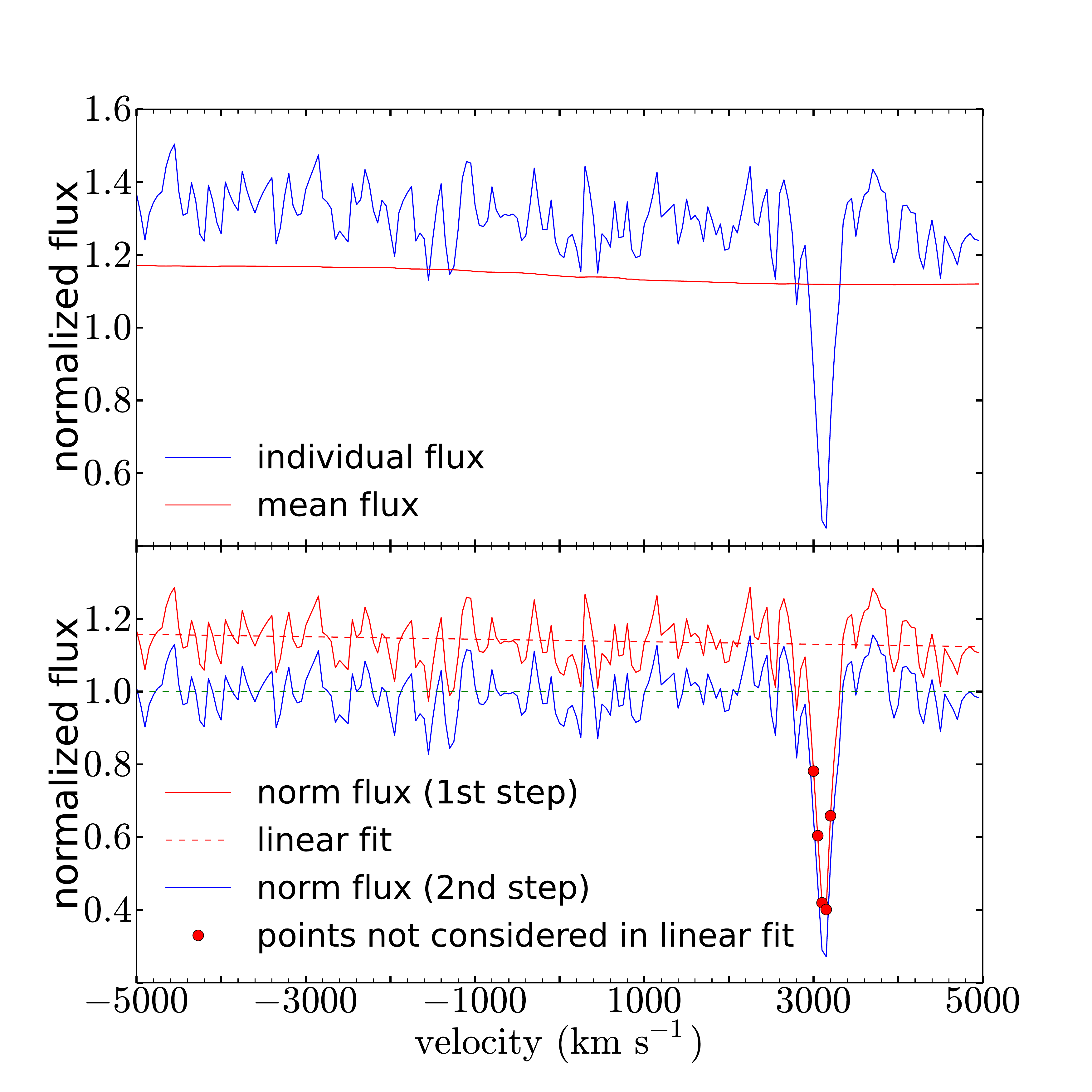}
	\caption{Examples illustrating the procedure explained in Section
 \ref{subsubs: Generating composite spectra}.
The left column shows
a case with a detected individual \mgii absorption system, and the right
column a case with no detectable associated  \mgii absorption system, but
with an unrelated metal absorption line. The top panels show the
normalized flux of the spectral region,
$f^{(r)}_{ik}/n_{j(k)}$ (solid blue line), and the normalized mean
spectrum $\overline{f}_{i}$ (dashed red line). The bottom panels show
the transmission $F^{(0)}_{ik}$ (thin, solid red line), the computed linear
fit $L_{ik}$ (thin, dashed red line), and the corrected transmission $F_{ik}$
(thick, solid blue line). In the bottom right panel, the points indicate pixels that
are excluded from the linear fit. A horizontal thick dashed green line at a
transmission of 1 is included for visual aid.}
	\label{fig: normalization example}
\end{figure*}

  The whole procedure is illustrated in figure
\ref{fig: normalization example} 
with a couple  of examples, one with an individually detected \mgii
absorption system and one without any individually detected \mgii
absorption but with a random metal absorption line. Note that the
contribution of these random metal absorption lines is later corrected
for (Section \ref{subs: Unbiasing the composite spectra}).

\subsection{Method 2: Variable smoothing} \label{subs: Spline fitting}

  As a second approach to determine the continuum, we use the method
described in \cite{York2005}, based on an iterative smoothing of the
spectrum with a variable smoothing length that is adjusted to decrease
in spectral regions of known quasar emission lines. Each quasar
spectrum is normalized by dividing both the flux and error by the
variable smoothing continuum model, $c_{ij}$:
\begin{equation}
	\label{eq: normalization 3 flux}
	F'_{ij}=f_{ij}/c_{ij} ~; \qquad E'_{ij}=e_{ij}/c_{ij} ~.
\end{equation}
In the same way as for the first method, the normalized spectra are
rebinned into the variable $v$ with binwidth of $50 \kms$. The final
composite spectrum is calculated using the same weights as in equation
(\ref{eq: weight}), following equation (\ref{eq: mean flux 2}), but
using the primed variables $F'_{ij}$ and $E'_{ij}$.

\subsection{Unbiasing the composite spectra}
 \label{subs: Unbiasing the composite spectra}

  After the stacking of all the normalized spectra of quasar-galaxy
pairs as a function of the velocity variable $v$ has been completed by
using either continuum fitting method, the mean value of the
transmission $\bar F_i$ that is obtained far from the expected \mgii
line (i.e., at large values of $\| v \|$) is close to unity but not
exactly so. The main reason for this is the presence of random metal
absorption lines (unrelated to the galaxy of the pair) that are detected
above a $3\sigma$ fluctuation and therefore excluded when fitting the
continuum. Other reasons may be affecting this mean
background of $\overline{F}_i$ related to systematic effects in the noise
distribution and the continuum fitting method. We eliminate this bias by
performing the same linear fit to the stacked spectrum as in Section
\ref{subsubs: Generating composite spectra}, using again the velocity
intervals $(-5000, -2000)$ and $(2000, 5000)\kms$ to measure two
average values of $\overline{F}_i$, and obtaining a linear fit based on two
points at the center of these intervals. The final normalized stacked
flux is found by dividing by this linear fit.

  Our results will actually be shown, for convenience, in terms of the
effective excess optical depth, defined according to
$\delta\tau_{{\rm e},i} = - \log \bar F_i$, in figures \ref{fig: stack1} to
\ref{fig: stack3}.

\subsection{Bootstrap errors} \label{subs: Bootstrap errors}

The errors computed from the known observational noise in the
observed quasar spectra that are stacked are actually a lower limit to
the true errors. In reality, the intrinsic variability of real quasar
continua and of the associated \mgii and other random metal absorption
lines imply the presence of additional errors that are not taken into
account and which are correlated among the pixels of the final stacks.
We therefore compute bootstrap errors, which are generally used in
our analysis and model fits in this paper.

To calculate these bootstrap errors, we use the BOSS plates as the regions
of the sky in which the sample is divided. Each quasar is tagged with the
plate number at which its best spectrum was observed. For DR7 there are
$\mathrm{N_{plates}}=1822$ plates
Pairs are counted as belonging to the plate that 
contains the quasar, irrespective of the galaxy position.

  Bootstrap errors are computed in the standard way, generating $N=1,000$
samples by randomly selecting $\mathrm{N_{plates}}$ among all the plates
with repetition, and then
recalculating the composite spectra using both methods. Bootstrap errors
are assigned to the final effective optical depth $\tau_{\rm e}$, at each
velocity bin and each impact parameter interval from the dispersion
found among the 1,000 random samples. These bootstrap errors are computed
for both continuum fitting methods.

We note that, specially at large impact parameter, some galaxies will be paired to quasars
in different plates. This implies the presence of
residual correlations among the bootstrap samples because of the common
galaxies in pairs belonging to different plates, but we believe this
effect is negligibly small because the most important error correlations
should arise from the quasar spectra.


\section{Model for the Absorption Profile}
\label{sec: absorption model}

  The usual analysis in the astronomical literature of individual \mgii
absorption lines is done by fitting with Voigt
profiles, with the equivalent width and the velocity dispersion as free
parameters. Whenever the observed absorption profile is not adequately
fitted in terms of the two Voigt profiles of the \mgii doublet, one can
include the presence of multiple cloud components with blended absorption
lines in order to improve the fit. Here, we are analyzing a stack of a
large number of \mgii absorption systems that may be mostly undetected
individually, but for which we can accurately predict the expected mean
position from the redshift of the galaxy near the quasar line of sight. The
effective optical depth in the stack, $\tau_{\rm e} = -\log(\bar F)$, should
in this case have a single symmetric component for each line in the
doublet, with a profile reflecting the velocity dispersion of \mgii
absorbing clouds and galaxies in halos, as well as the large-scale halo
correlation in redshift space for large impact parameters. This
cross-correlation function can be
modeled in terms of the Halo Occupation Distribution formalism 
\citep[e.g.,][]{Gauthier2008,Zhu&Menard2013a}, but the density profile
of \mgii clouds in halos does not have to follow that of
galaxies, and it will generally depend on complex physics of galaxy
winds and gas accretion in the circumgalactic medium. For
simplicity, we shall assume in this section a model with a Gaussian
velocity distribution and a power-law form for the projected correlation
function, as an approximation to the generally complex form of the
galaxy-absorber cross-correlation function. In the next section, we
shall use a more accurate form of the correlation function obtained
from halo simulations to determine the amplitude of our measured
cross-correlation.

Our model has four free parameters. The first three are the central
effective optical depth $\tau_0$ of the strongest line in the doublet at
a conventional normalization value of the impact parameter $r_{p0}$ (set
here to 1 Mpc), the power-law slope $\alpha$ of the projected
cross-correlation,
and the mean equivalent width ratio $q$ (where $q=1/2$ if all the
absorption lines were unsaturated, and $q=1$ in the saturated case).
We consider a variation of the velocity dispersion $\sigma$ with radius,
taking into account that in the limit of large radius, the Hubble
expansion should lead to a linear increase of the effective dispersion.
The fitted profile of the excess effective optical depth,
$\delta\tau_{\rm e}(b,v)$ (as defined in equation \ref{eq: deltatau}), is
\begin{displaymath}
	\delta{\tau}_{\rm e}(r_{\rm p},v) =
 \frac{\tau_{0}}{\sqrt{1+\left(xHr_{\rm p}/\sigma_{0}\right)^{2}}}
 \left( \frac{r_{\rm p}}{r_{\rm p0}}\right)^{-\alpha}\times
\end{displaymath}
\begin{equation}
	\label{eq: absorption model}
 \times\left[{\mathrm e}^{-\left(v-v_{1}\right)^{2}/2\sigma^{2}}
 + q\, {\mathrm e}^{-\left(v-v_{2}\right)^{2}/2\sigma^{2}}\right] ~,
\end{equation}
where
\begin{equation}
 	\sigma^{2}=\sigma_{0}^{2}+\left(xHr_{\rm p}\right)^{2} ~.
\end{equation}
The fourth model parameter is therefore the dimensionless constant $x$,
which is the scale at which the velocity dispersion starts to increase,
compared to $\sigma_0/H$, with $H(z=0.55) = 91.7$. 
In principle, the central velocity dispersion
$\sigma_{0}$ should also be left as a free parameter, but in this paper
we have fixed it to $250\kms$. The reason is that if this condition is
relaxed, the obtained fits have a large degeneracy in $\sigma_{0}$ and
$x$ and they are largely dominated by an excess of absorption that is
found at the largest impact parameters that we do not fully understand,
as we shall see below. The value of $250 \kms$ allows us to fit well
the width of the doublet line at small impact parameter, where it is
resolved. Note that the parameter $q$ is assumed to be
independent of impact parameter, even though it may generally depend
on it (the mean saturation should decrease with impact parameter
if the mean absorber equivalent width also decreases, as suggested by
observations; e.g.,\cite{Gauthier2009,Chen2010a}). Note also that the
instrumental Point Spread Function (PSF) \citep{Smee2013} and finite
spectral bin size is effectively included in the value of the velocity
dispersion in our model. A more detailed modeling of these effects is
neglected here for simplicity. The origin of the velocity coordinate $v$
is chosen by convention as the central position of the \mgii doublet for
an unsaturated line ($q=0.5$) at the redshift of the galaxy. Under this
convention, $v_1=-256.05 \kms$ and $v_2=513.28 \kms$.

The fitting is performed by $\chi^{2}$ minimization, computed including
all the pixels of the stacked spectra in the central interval
$|v|<2\,000 \kms$. Each pixel is weighted according to the bootstrap
error measured in that pixel as described in Section
\ref{subs: Bootstrap errors}, but without considering the
cross-correlations of the errors between pixels. In practice,
the bootstrap errorbars in different pixels of the stacked spectra are
nearly equal. The spectrum outside the interval $|v|<2\,000\kms$, which
is used for deriving the continuum fit in Method 1, is not considered
here for the fit. We note also that for a real Gaussian profile for each
component of the \mgii doublet, the continuum fitting of Method 1 implies
that we formally need to subtract a small constant from the double
Gaussian in equation (\ref{eq: absorption model}), equal to the
integrated value of the model absorption over the interval
$v\in (2000, 5000) \kms$ used for determining the continuum, but we
neglect this effect here.

We use a Monte Carlo Markov Chain method to perform the $\chi^2$
minimization to the five-parameter model fit. The errors of the
parameters are also obtained by repeating the model fit with bootstrap
realizations. The average integrated equivalent width as a function of
impact parameter can be obtained directly by integrating the effective
optical depth over the interval used to fit the model,
\begin{equation}
	\label{eq: EW direct}
  W_{\rm e}(r_{\rm p}) =\frac{\lambda_{\rm \mgii}}{c} \int_{-2000\, {\rm \kms}}^{2000\, {\rm \kms}}\,
  \delta{\tau}_{e}(r_{p},v)\, dv ~.
\end{equation}
The fitted model also predicts a mean equivalent width,
\begin{displaymath}
  W_e(r_{p}) = \frac{\lambda_{\rm \mgii}}{c} \sqrt{2\pi}\, \tau_{0}\sigma_{0} (1+q)\,
 \left(\frac{r_{p}}{r_{p0}}\right)^{-\alpha}=
\end{displaymath}
\begin{equation}
	\label{eq: EW model}
 = W_{e0} \left(\frac{r_{p}}{r_{p0}}\right)^{-\alpha} ~,
\end{equation}
although this value is for the absorption over the whole velocity
range, not restricted to the interval $-2000 \kms < v < 2000 \kms$.

\section{Results} \label{sec: Results}

  The results of the stacked absorption profiles are obtained for a
total of 17 impact parameter intervals, measured in proper units at the
redshift of the galaxy. The first interval is for
$r_{\rm p}<50\, {\rm kpc}$, and the other 16 intervals are
$2^{(i-1)/2} < (r_{p}/50\, {\rm kpc}) < 2^{i/2}$, for $i=1$ to 16, up to
a maximum impact parameter of $12.8$ Mpc. The stacked profiles are shown
as the effective optical depth, $\tau_{\rm e}=-\log(\bar F)$, in figures
\ref{fig: stack1}, \ref{fig: stack2} and \ref{fig: stack3}. Results are
presented for our two continuum fitting methods, the mean subtraction
method (thick, solid blue line) and the variable smoothing method (thin,
solid red line). The errorbars plotted on the left side are the rms
value of the bootstrap error of $\tau_{\rm e}$ in one pixel, which has
little variation among different pixels in each stacked spectrum. The
results of the fitted model parameters and their bootstrap errors
computed by repeating the fit for bootstrap realizations of the stacked
profiles, are given in table
\ref{ta: absorption model values} (the cross-correlations of the
parameter errors are omitted). 


\begin{table}
	\centering
	\begin{tabular}{c|c|c}
						&	mean subtraction	& variable smoothing \\ \hline
	$\tau_{0}$			& $0.0060\pm0.0001$	& $0.0034\pm0.0001$	\\
	$\alpha$ 				& $0.70\pm0.01$		& $0.88\pm0.01$		\\
	$\sigma_{0}\, [\kms ]$	& $250$				& $250$				\\
	$x$					& $1.35\pm0.06$		& $0.46\pm0.08$		\\
	$q$	 				& $0.65\pm0.03$		& $0.66\pm0.03$		\\
	$W_{e0} [ \kms ]$		& $6.27\pm0.10$		& $3.58\pm0.08$		\\
	\end{tabular}
	\caption{Best-fit values for the fitted parameters of the model
described in Section 4 and shown as the lines in Figure 9.
{\it First column:} Results using the mean subtraction method
(see Section \ref{subs: Mean subtraction}).
{\it Second column:} Results using the variable smoothing method
(see Section \ref{subs: Spline fitting}).
Four independent parameters are fitted, and $W_{\rm e0}$ is related
to the other parameters according to equation (\ref{eq: EW model}). As explained in section
\ref{sec: absorption model}, $\sigma_{0}$ is fixed at $250\kms$.
Errors are computed from repeating the fits on bootstrap realizations of
the stacked profiles. }
	\label{ta: absorption model values}
\end{table}
	
\begin{figure*}
	\includegraphics[width=\textwidth]{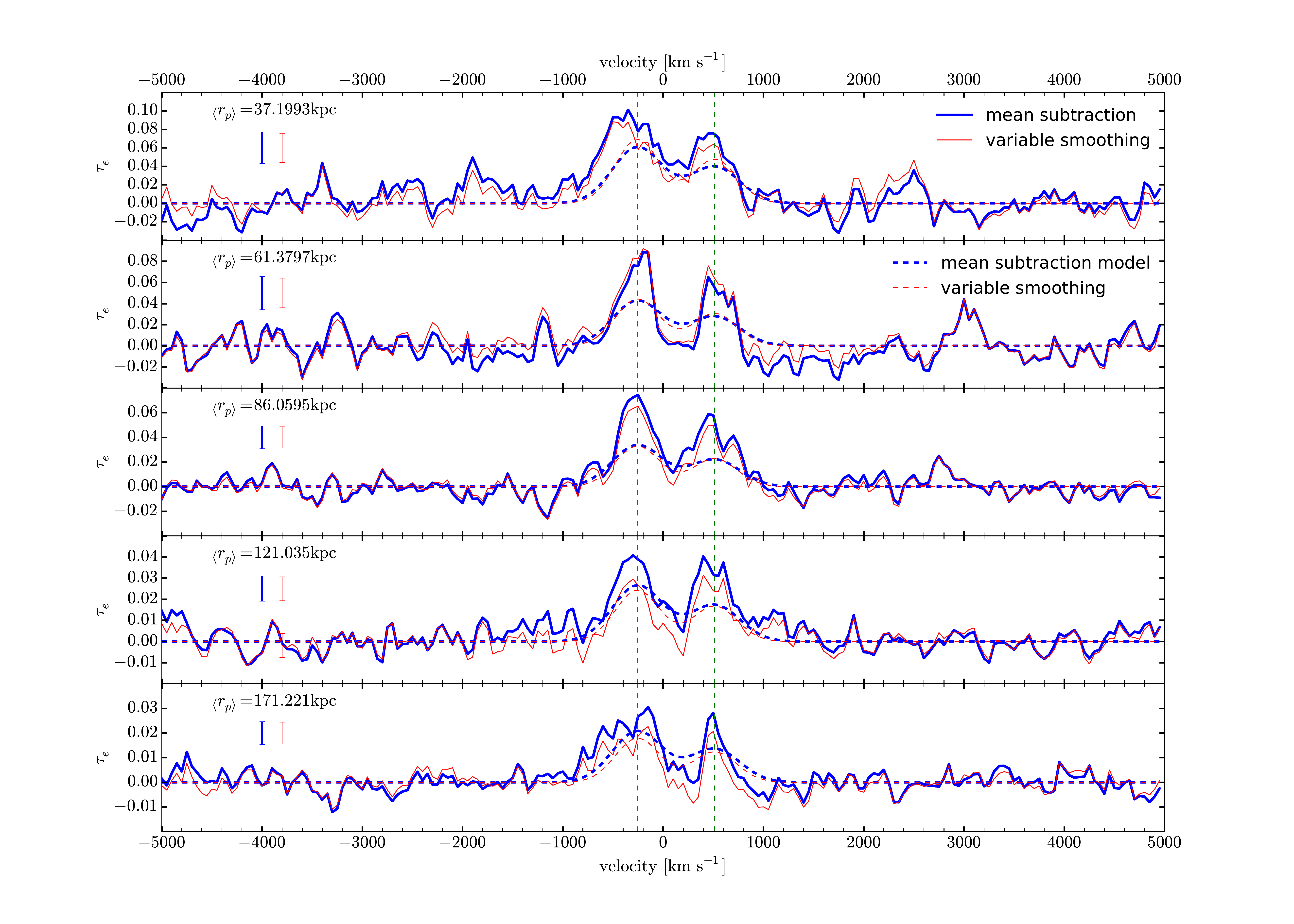}
	\caption{From top to bottom, composite spectra for increasing
impact parameter intervals (in proper kpc). The effective optical depth
is shown against velocity for the mean subtraction method (
thick, solid blue line) and the variable smoothing method (thin, solid red line). The rms value
of the bootstrap error in individual pixels is shown by the errorbars
on the left. The thick,
dashed blue line and the thin, dashed red line are the best fit model 
(equation \ref{eq: absorption model})
for the mean subtraction and variable smoothing methods respectively. A single set of parameters are fitted
to all the 17 regions. Figures \ref{fig: stack2} and \ref{fig: stack3}
show the spectra for the remaining impact parameter intervals. The stacks show a mean
absorption profile for the presence of the \mgii doublet line at the expected position.
For visual gidance, vertical, dashed green lines mark the predicted position of the \mgii doublet.}
	\label{fig: stack1}
\end{figure*}
\begin{figure*}
	\includegraphics[width=\textwidth]{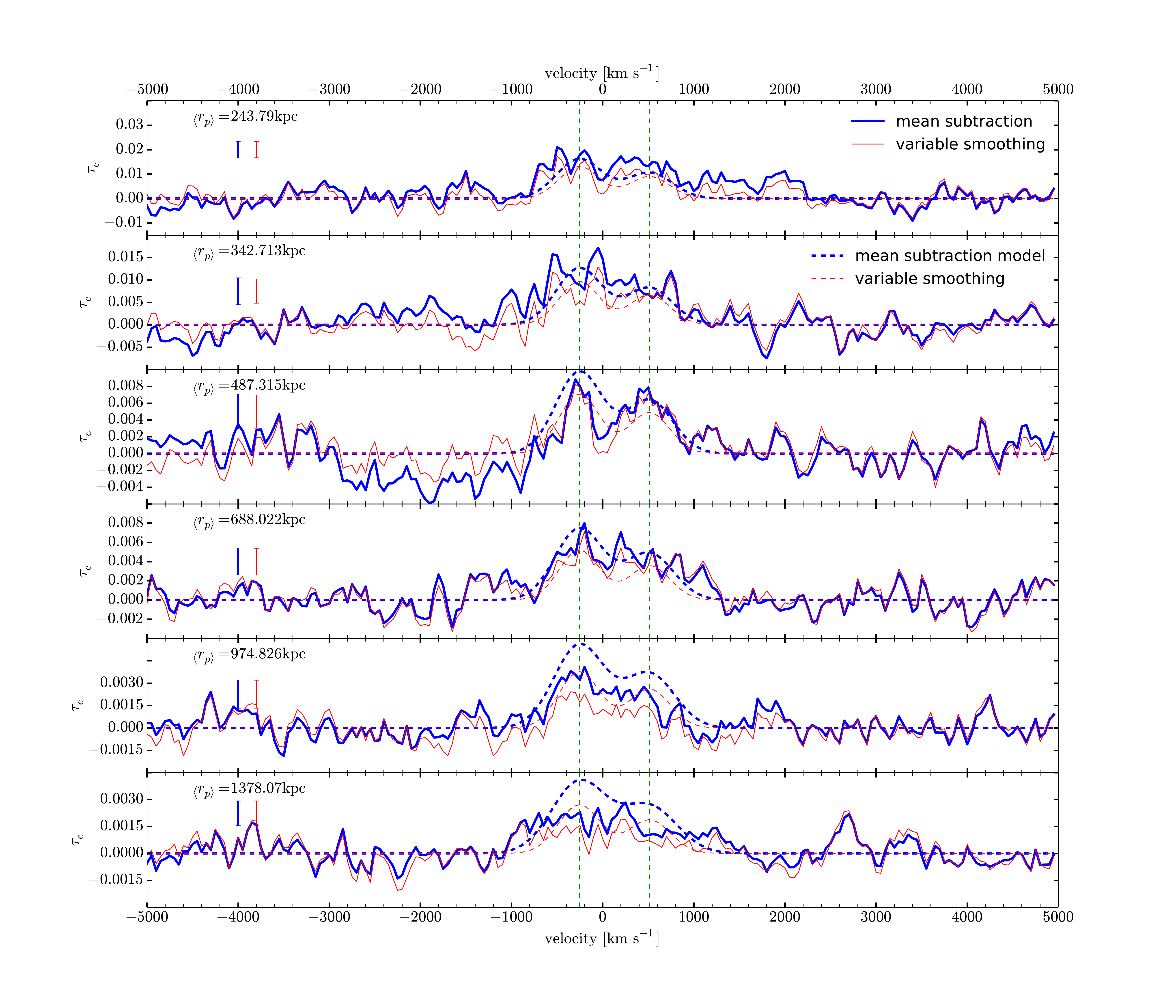}
	\caption{Continuation  of figure \ref{fig: stack1}.}
	\label{fig: stack2}
\end{figure*}
\begin{figure*}
	\includegraphics[width=\textwidth]{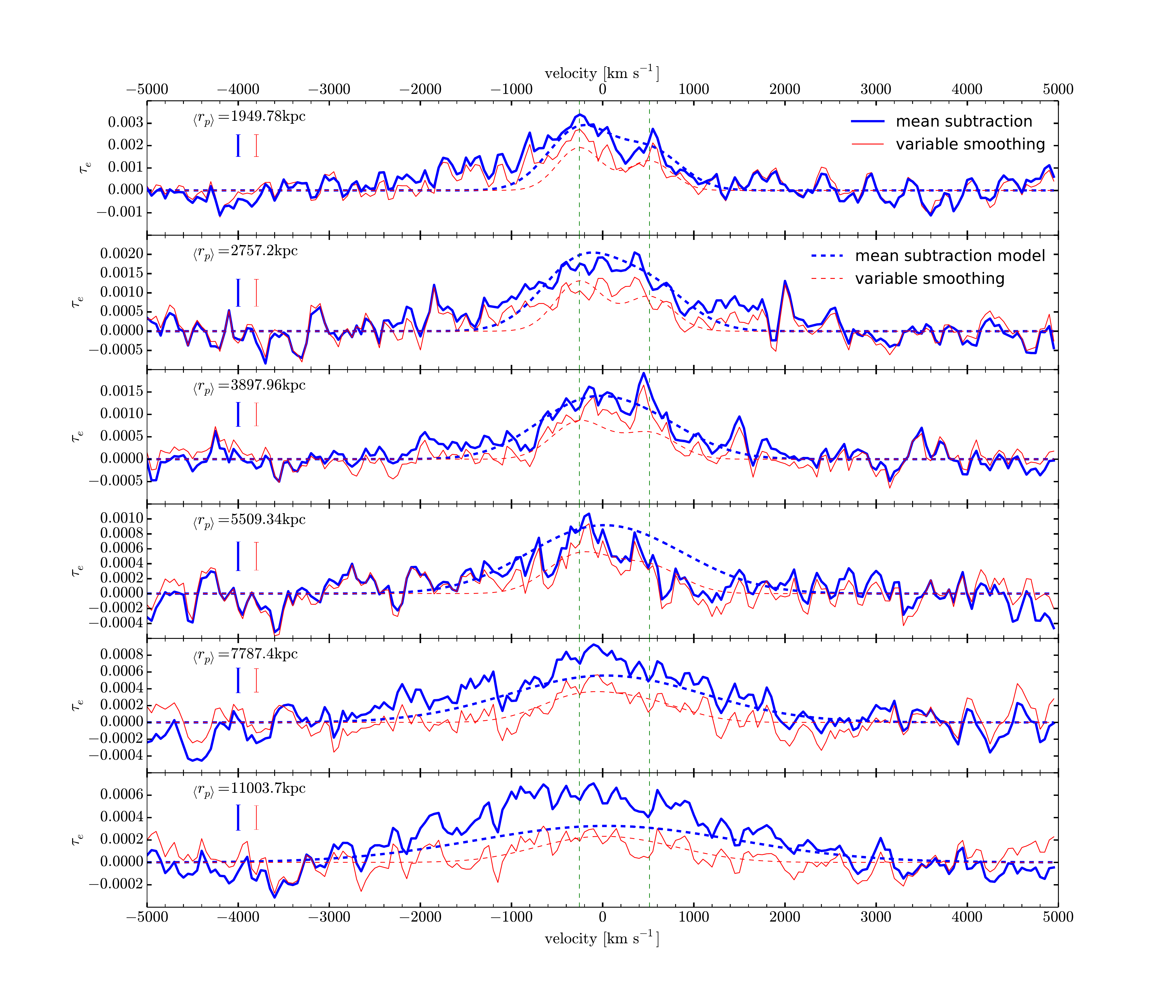}
	\caption{Continuation  of figure \ref{fig: stack1}.}
	\label{fig: stack3}
\end{figure*}

  The stacks in figures \ref{fig: stack1},
\ref{fig: stack2} and \ref{fig: stack3} show the mean absorption
profile of the \mgii doublet line, clearly seen at the expected
positions at small impact parameters. The
amplitude of the random pixel-to-pixel variations outside of the central
absorption feature is generally consistent with the computed bootstrap
errorbars. The expected doublet feature of the \mgii line is generally
resolved at $r_{\rm p} \la 200$ kpc, and
is smoothed out at larger impact parameter by the velocity dispersion
of the absorbers, which should increase linearly with $r_p$ in the limit
of large radius because of Hubble expansion.
The fact that our parameter $x$ is close to unity supports this
interpretation, although we note that the precise expected theoretical
value of $x$ in the linear regime is not one because of redshift
distortions. We do not analyze this issue further in this
paper because the maximum impact parameter that we analyze is not yet
in the linear regime, and a more detailed model would
be necessary for the velocity distribution of absorbers as a function
of impact parameter. We mention here that we initially considered a
simpler model with $x=0$ and $\sigma_{0}$ as free parameter, but this
choice gave a substantially worse fit and resulted in a high
value of the velocity dispersion because of its increase with
radius. As mentioned previously, fits leaving both $\sigma_{0}$ and
$x_0$ as free parameters lead to a substantial degeneracy and to
solutions with very large values of $\sigma_0$, driven by the excess
of absorption that is seen in the last $r_p$ bin for the mean
subtraction method. We have not found an explanation for this excess
in the last bin, which is only marginally consistent
(at the $\sim 3-\sigma$ level, as shown below in figure 8) with our fit.

 A value $q\simeq 0.65$ is obtained for the line ratio of the \mgii
doublet, consistent with a mixture of saturated and unsaturated lines.
It has been previously reported that the mean equivalent width of the \mgii
lines decreases with impact parameter \citep[and
references therein]{Gauthier2009}. This result may imply that absorbers are less
saturated at larger impact parameters, and should therefore have a
decreasing value of $q$, although this interpretation depends on whether the internal
velocity dispersion of the absorbing clouds varies with impact
parameter (note that this internal velocity dispersion is much smaller
than the velocity dispersion of absorbing components around their host
galaxies). Our model assumes a constant value of $q$ for simplicity.

\begin{figure*}
	\centering
	\includegraphics[width=\textwidth]{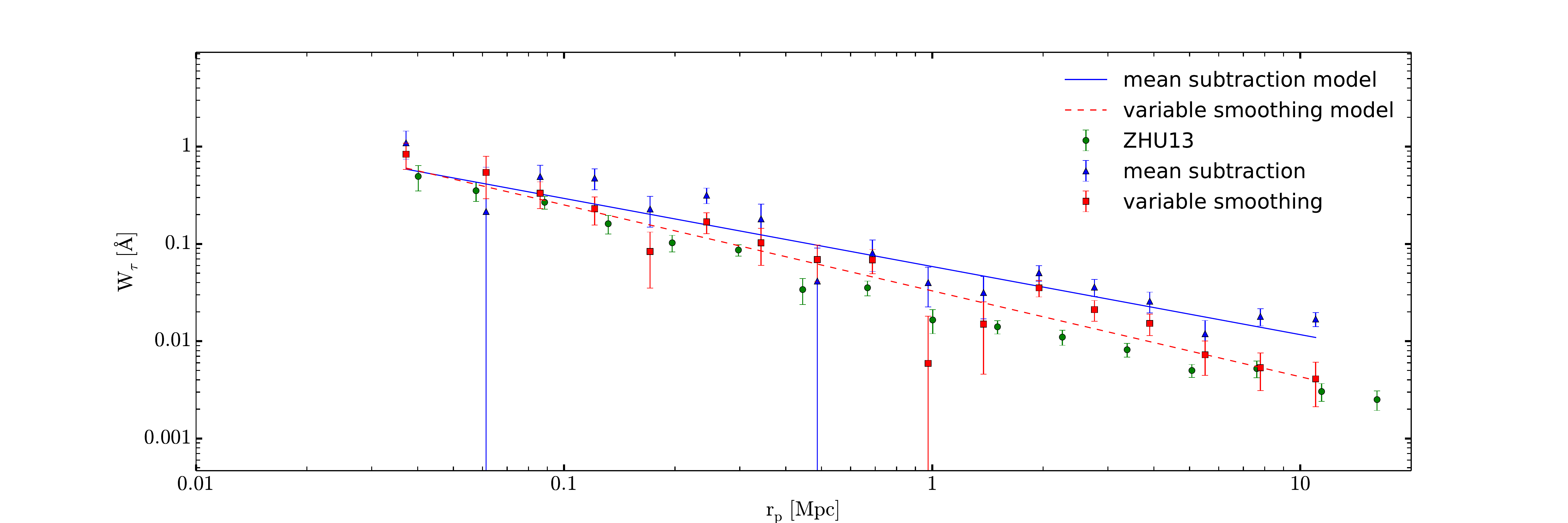}
	\caption{Measured rest-frame mean equivalent width $W_{\rm e}$
of the \mgii doublet, versus proper
impact parameter. Blue triangles are obtained from the mean subtraction
continuum method, and red squares use the variable smoothing method.
Errorbars have been obtained by the bootstrap method. Lines are the
best-fit power-law model to the data for both methods. Green circles are
the results of \protect\cite{Zhu2013}. Note that the results for the mean subtraction 
method are systematically higher than the rest. This is explored in further
detail in the appendix \ref{sec: Test}. }
	\label{fig: ew_plot}
\end{figure*}

  The mean equivalent widths obtained with our two methods, by directly
integrating the effective optical depth in the stacked spectra as in
equation (\ref{eq: EW direct}), are shown in figure \ref{fig: ew_plot}
as blue triangles for the mean subtraction method, and red squares for
the variable smoothing method. These values and their bootstrap errors are
also given in table \ref{ta: ew_plot}, together with the number of
galaxy-quasar pairs that contribute to the stacked spectrum at each
impact parameter bin.
Note that this mean equivalent width is for
the sum of the two lines in the \mgii doublet. There is a systematic
difference in the mean equivalent width obtained with the two methods;
the variable smoothing method yields a systematically
smaller equivalent width compared to the mean subtraction method, and
the discrepancy increases with impact parameter, reaching more than a
factor $2$ at $r_{\rm p}= 10$ Mpc. As mentioned previously in Section
\ref{sec: Stacking procedure}, the reason for this difference is that
the spectral region where \mgii absorption is expected is used to
determine the continuum in the variable smoothing method. The presence
of weak lines that
remain undetected in individual spectra causes the continuum to be
underestimated in a way that depends on the signal-to-noise ratio and the
equivalent width of the undetected line in a complex manner. This
systematic underestimate of the continuum causes the underestimate of
the mean equivalent width. Appendix \ref{sec: Test} presents
quantitative tests demonstrating the presence of this systematic
error of the variable smoothing method, and shows also that the result
obtained with the mean subtraction method, which does not use the
stacked spectrum in the region of the \mgii absorption to determine the
continuum, is free of any similar systematic effect to the extent that
we are able to discern.

  The green circles with errorbars in figure \ref{fig: ew_plot} show
the results of \cite{Zhu2013}, who have used a sample of galaxies and
quasar spectra similar to ours to infer the same mean equivalent
width as a function of impact parameter. Their result is systematically
below ours, roughly by a factor $\sim 2$ at all impact parameters,
and is lower even compared to our variable smoothing method. We believe the
reason is again due to the systematic underestimate of the continuum. The
continuum fitting method used by \cite{Zhu2013} also uses the observed
spectra in the region where \mgii absorption is expected in a rather
complex way that is described in \cite{Zhu&Menard2013b}, and the systematic error that
this induces is difficult to predict but may in principle explain why it
produces a systematically low estimate of the equivalent width. 
Note that in principle this underestimation will only affect the individually undetected systems. Individually undetected systems cannot be distinguished from the noise and will thus be fitted away by the continuum fitter. Individually detected systems will, in principle, not suffer from this effect.
The errorbars of \cite{Zhu2013} are also smaller than ours, since their
continuum fitting can better remove any features of the quasar spectrum
that are superposed with the \mgii absorption lines, at the cost of
introducing a systematic bias in the continuum estimate.

\begin{table*}
	\centering
	\begin{tabular}{c|c|c|c|c|c}
	\multirow{2}{*}{$r_{p}$} & \multicolumn{2}{c|}{mean subtraction} & \multicolumn{2}{c|}{variable smoothing} & \multirow{3}{*}{$N_{\rm pairs}$ }\\
 &  $<W>$ & $\sigma\left(<W>\right)$ & $<W>$ & $\sigma\left(<W>\right)$ & \\ 
 $\left[ {\rm kpc} \right]$ & [\angs] & [\angs] & [\angs] & [\angs] & \\	\hline
$\left(0,50\right]$ 				& 1.10	& 0.35	& 0.84	& 0.25	& 76			\\
$\left(50,70.71\right]$ 			& 0.22	& 0.40	& 0.54	& 0.25	& 94			\\
$\left(70.71,100\right]$ 	    		& 0.49	& 0.15	& 0.33	& 0.10	& 204		\\
$\left(100,141.42\right]$		& 0.48	& 0.12	& 0.230	& 0.074	& 423		\\
$\left(141.42,200\right]$		& 0.229	& 0.080	& 0.084	& 0.049	& 758		\\
$\left(200,282.84\right]$		& 0.317	& 0.058	& 0.168	& 0.041	& 1396		\\
$\left(282.84,400\right]$		& 0.181	& 0.076	& 0.102	& 0.042	& 2621		\\
$\left(400,565.69\right]$		& 0.042	& 0.049	& 0.069	& 0.028	& 4904		\\
$\left(565.69,800\right]$		& 0.081	& 0.029	& 0.069	& 0.019	& 9576		\\
$\left(800,1131.37\right]$		& 0.040	& 0.018	& 0.006	& 0.012	& 19166		\\
$\left(1131.37,1600\right]$		& 0.032	& 0.015	& 0.015	& 0.010	& 38922		\\
$\left(1600,2262.74\right]$	       	& 0.0506	& 0.0093	& 0.0354	& 0.0068	& 75567		\\
$\left(2262.74,3200\right]$	    	& 0.0360	& 0.0072	& 0.0211	& 0.0051	& 148769		\\
$\left(3200,4525.48\right]$        	& 0.0258	& 0.0062	& 0.0152	& 0.0038	& 290340		\\
$\left(4525.48,6400\right]$      	& 0.0119	& 0.0044	& 0.0073	& 0.0028	& 559840		\\
$\left(6400,9050.97\right]$     	& 0.0180	& 0.0036	& 0.0053	& 0.0022	& 1062482	\\
$\left(9050.97,12800\right]$       	& 0.0169	& 0.0028	& 0.0041	& 0.0020	& 1961450	\\ 
\multicolumn{6}{c}{ }\\
	\end{tabular}
	\caption{Results on the mean equivalent width and errors shown in figure
\ref{fig: ew_plot},
presented here as a table. From  left to right, impact
parameter interval, mean rest-frame \mgii equivalent width and its
bootstrap error for the mean subtraction and
variable smoothing methods, and number of QSO-galaxy pairs used in the
interval. The mean rest-frame equivalent widhts are the sum of both
lines in the doublet.}
	\label{ta: ew_plot}
\end{table*}

\section{Discussion} \label{sec: Discussion}

\subsection{Relation of the mean equivalent width to the bias factor
of \mgii absorption systems} \label{subs: biasfactor}

  Our measurement of the cross-correlation of \mgii absorption systems
and galaxies in the CMASS catalog of BOSS clearly reflects properties
of the spatial distribution of these two objects. In the limit of
large scales, when the fluctuations are in the linear regime, any
population of objects that traces the large-scale mass perturbations
is characterized by its bias factor
and the autocorrelation in real space is equal to the correlation
function of the mass times the square of the bias factor, with
redshift distortions added in redshift space
 \citep{Kaiser1987,Cole&Kaiser1989}. The
cross-correlation of two classes of objects is, in the same large-scale
limit, equal to the mass correlation function times the product of the
two bias factors. On small, non-linear scales, the correlations are more
complex and they depend on other physics that determine the distribution
of galaxies and \mgii absorbers in relation to dark matter halos.

  Our stacked spectra measure the mean excess of the effective optical
depth as a function of impact parameter $r_{p}$ and velocity separation $v$
from a galaxy. This quantity is related to the mean \mgii absorption perturbation,
$\delta_{\mathrm{Mg}}(r_{\rm p},v)$, as
\begin{equation}
	\tau_{\rm e}(r_{\rm p},v)=\tau_{\rm e0}\left(1+\delta_{\mathrm{Mg}}(r_{\rm p},v)\right)
\end{equation}
where $\tau_{\rm e0}$ is the mean \mgii effective optical depth, defined in
equation \ref{eq: tau_e0}. This perturbation is equal to the
cross-correlation function of \mgii absorbers and galaxies, convolved
with the mean doublet absorption profile of a \mgii system, and is the
function that is measured in our stacking results in figures
\ref{fig: stack1} to \ref{fig: stack2}. In this work, our interest is
focused on the projected correlation function, related to the integrated
absorption
$W_e(r_{\rm p})$ in equation (\ref{eq: EW direct}). The projected
cross-correlation is not affected by redshift distortions and by the
complications arising from the convolution with the mean doublet line
profile and the spectrograph resolution.

  Here, we shall make two approximations to physically interpret our
measurement of $W_{\rm e}(r_{\rm p})$: first, we neglect the effect of
the finite integrating range $\pm 2000 \kms$ that we have used, ignoring
the difference from the true projected correlation that is obtained by
integrating to infinity. This approximation is likely not very good for
the largest impact parameters we use; we discuss this further below.
Second, we assume that the cross-correlation of \mgii systems and CMASS
galaxies is the same as the auto-correlation of CMASS galaxies times the
ratio of bias factors $b_{\mathrm{Mg}}/b_{\rm g}$ of the two
types of objects. In other words, we assume the linear relation can
be extended into the non-linear regime as far as the ratio of the
cross-correlation to the auto-correlation is concerned.

  This second assumption can be justified from observations of the
correlations of galaxies of different luminosity. \cite{Zehavi2011}
measured the projected correlation of galaxies in the DR7 catalogue 
in different luminosity ranges, and, to a good approximation, in
the impact parameter range of our interest, the result is a fixed shape
times the variable bias factor, as seen for example in their figure 6.
The shape of the galaxy correlation can be interpreted as arising
from the correlation among virialized halos and the distribution of
galaxies within each halo \citep[e.g.][]{Zheng2005}. This shape does
vary slightly with luminosity, but the most important variation is
the normalization determined by the bias factor. There is a greater
variation of the shape of the projected correlation with galaxy color
\citep[see figure 21 in][]{Zehavi2011}. In addition, the projected
cross-correlation of galaxies of different color is not exactly equal to
the geometric mean of the projected auto-correlations of the two types
of galaxies (see their figure 15). Our assumption can only be considered
as a first approximation that will need to be tested in the future, but
it allows us to obtain a bias factor for \mgii absorption systems
assuming that they behave in a similar way as galaxies in the CMASS
catalog.

  These assumptions lead to the relation
\begin{displaymath}
 W_{\rm e}(r_{\rm p}) = \frac{\tau_{\rm e0}\lambda_{\rm \mgii}}{c} \int {\rm d}v \, \delta_{\mathrm{Mg}}(r_{\rm p},v)=
\end{displaymath}
\begin{equation}
	\label{eq: delta MgII}
=\frac{\lambda_{\rm \mgii}}{c}
 {\tau_{\rm e0}\, H(z)\over 1+z}\, \frac{b_{\mathrm{Mg}}}{b_{\rm g}}\, w_{\rm gg}(r_{\rm p}) ~,
\end{equation}
where $w_{\rm gg}(r_{\rm p})$ is the projected galaxy correlation function,
$b_{\rm g}$ is the galaxy bias factor and $b_{\mathrm{Mg}}$ is the mean bias
factor of \mgii absorption systems, weighted in proportion to their
equivalent width. We have used ${\rm d}v= H(z)/(1+z)\, {\rm d}x$, where
${\rm d}x$ is the comoving space coordinate that is integrated to obtain
the projected galaxy correlation function, and $z$ is the mean redshift
of the galaxies and associated \mgii absorption systems. This relation
allows us to infer the bias factor of \mgii systems empirically, using
only the directly measured projected galaxy correlation. Its validity is
strictly valid in the limit of large scales, but, as we shall see below,
the ratio $W_{\rm e}(r_{\rm p})/w_{\rm gg}(r_{\rm p})$ is roughly
constant, making our assumption plausible as a first approximation.

\subsection{Mean absorption from \mgii systems}

  The value of $\tau_{\rm e0}$, representing the average absorption from
the population of \mgii absorbers, must
be independently known before we can
use the measured mean excess of \mgii absorption around galaxies to
infer the bias factor of \mgii systems with equation (\ref{eq: delta MgII}).
This parameter can be estimated from
equation (\ref{eq: tau_e0}) using models of the equivalent width
distribution that fit the observational data.

\begin{figure}
	\includegraphics[width=0.5\textwidth]{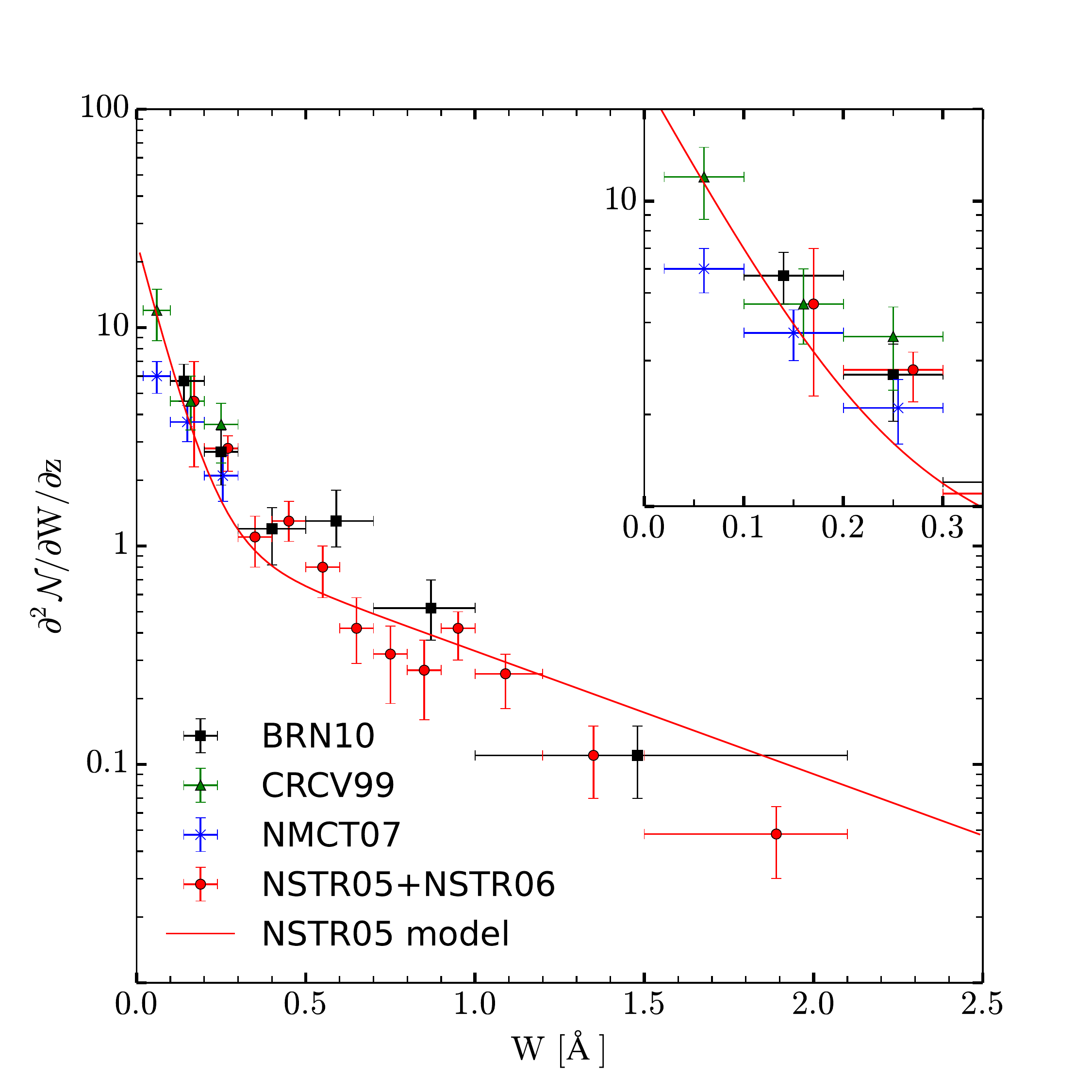}
	\caption{Rest-frame equivalent width distribution of \mgii
absorption systems. Data points are from \protect\cite{Bernet2010}
(black squares), \protect\cite{Churchill1999} (green triangles),
\protect\cite{Narayanan2007} (blue crosses) and \protect\cite{Nestor2005,Nestor2006}
(red circles). The overplotted solid line is the double exponential fit
of \protect\cite{Nestor2005} (see text). Top right pannel is a zoomed view of
the weakest absorption systems.}
	\label{fig: MgII distribution}
\end{figure}

  We use the double exponential model of \cite{Nestor2005},
\begin{equation}
 \label{eq: dexp}
	\frac{\partial^{2} \mathcal{N}}{\partial W\partial z} =
 \frac{N^{*}_{\rm wk}}{W^{*}_{\rm wk}}\exp^{-W/W^{*}_{\rm wk}} +
 \frac{N^{*}_{\rm str}}{W^{*}_{\rm str}}\exp^{-W/W^{*}_{\rm str}} ~,
\end{equation}
where $N_{\rm str}^{*}$ and $N_{\rm wk}^{*}$ are the number of absorbers per unit
of redshift in the strong and weak population, and $W_{\rm str}^{*}$ and
$W_{\rm wk}^{*}$ are the characteristic rest-frame equivalent widths of the
two exponential distributions.
This model was fitted by \cite{Nestor2005} to their data, using mocks
to correct for incompleteness at low equivalent widths. More recently,
a compilation of high-resolution data was shown by \cite{Bernet2010}
in their figure 5, reaching down to lower equivalent widths. We
include these observational results in figure
\ref{fig: MgII distribution}, overplotting the fit that was obtained
by \cite{Nestor2005}, which has the following parameters:
$N^{*}_{\rm wk}=1.71\pm0.02$, $W^{*}_{\rm wk}=0.072\pm0.001$\angs,
$N^{*}_{\rm str}=0.932\pm0.011$ and $W^{*}_{\rm str}=0.771\pm0.014$\angs.
The observations are well reproduced by this fit, which
we therefore use to compute $\tau_{\rm e0}$.

  Unfortunately, the sample of absorbers of \cite{Nestor2005} is
somewhat heterogeneous, and the main uncertainty we encounter in using
it to compute $\tau_{\rm e0}$ is due to the redshift evolution. A fit to
this evolution was determined by \cite{Nestor2005}, where the parameters
of the exponential model vary as $W^{*} \propto (1+z)^{0.634\pm 0.097}$
and $N^{*} \propto (1+z)^{0.226\pm 0.170}$, both for the weak and strong
populations. We infer from their model fits and the mean value of
$N^{*}$ that the mean redshift of their sample is $z\simeq 1.1$ and we
use these relations to convert the product $N^{*} W^{*}$ to the mean
redshift of the CMASS galaxy catalog, $z\simeq 0.55$. We find
$N_{\rm str}^{*} W_{\rm str}^{*} = 0.55$\angs and
$N_{\rm wk}^{*} W_{\rm wk}^{*} = 0.095$\angs, with an error that is close to
10\%, although it is poorly defined
because the errors in the redshift evolution of 
$N_{\rm str,wk}^{*}$ and $W_{\rm str,wk}^{*}$ should be correlated, and this information (and the
exact redshift distribution of the absorbers) was not provided in
\cite{Nestor2005}.

  From equation \ref{eq: tau_e0}, we derive the value of $\tau_{\rm e0}$ in
the double exponential model of equation \ref{eq: dexp},
\begin{equation}
	\tau_{\rm e0} = \frac{1+z}{\lambda_{\rm{\mgii}}}\, (1+\bar q)\,
 \left(N_{\rm wk}^{*}W_{\rm wk}^{*}+N_{\rm str}^{*}W_{\rm str}^{*} \right) ~.
\end{equation}
which yields a value $\tau_{\rm e0}(z=0.55)=5.0\times 10^{-4}$, with an
error of about 10\% but which is subject to uncertainties owing to
the redshift evolution, the accuracy of the fit to the equivalent
width distribution, and the value of $\bar q$. Our results will be given
in terms of $\tau_{\rm e0}$ without including its error, with the
understanding that this quantity will need to be better determined in
the future from studies of the field population of \mgii absorbers.

\subsection{Derivation of the bias factor of \mgii systems}
 \label{subs: Derivation of the bias factor of MgII systems}

\begin{figure*}
	\includegraphics[width=\textwidth]{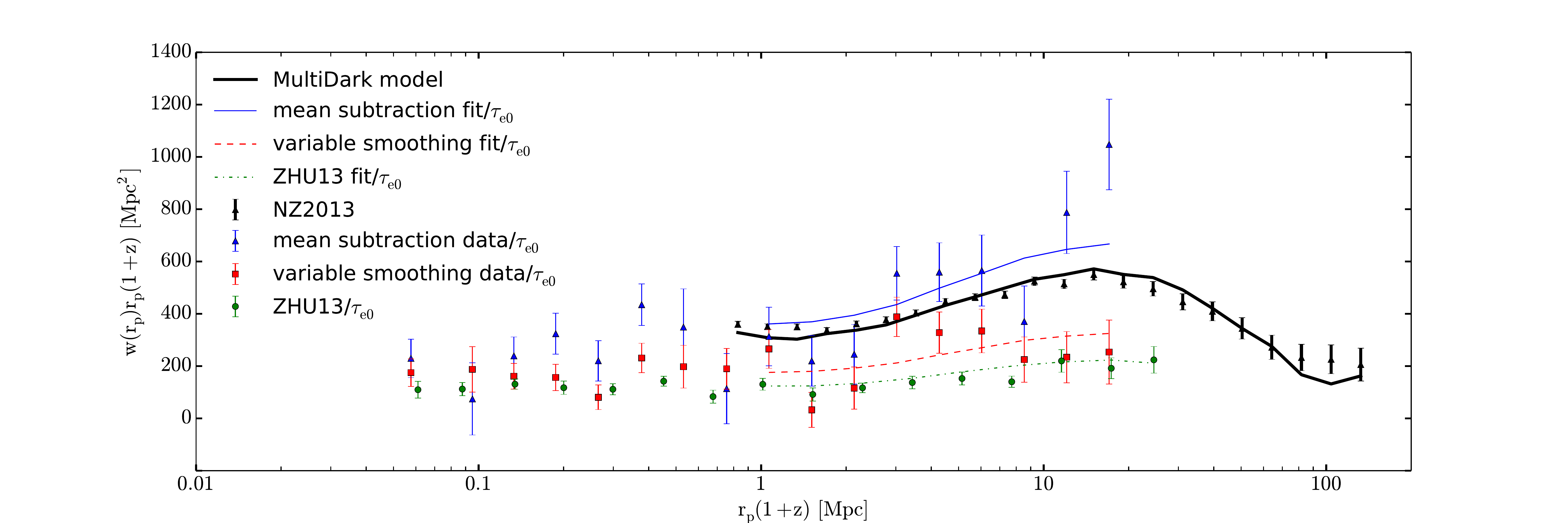}
	\caption{Projected correlation functions multiplied by the comoving impact
parameter as a function of the comoving impact parameter $r_{\rm p}(1+z)$.
Thick black triangles with errorbars show the auto-correlation of CMASS galaxies
from \protect\cite{Nuza2013}. Blue triangles, red squares and green circles are the mean equivalent width
$W_{\rm e}(r_{\rm p})$ times the factor $(1+z)/H(z)/\tau_{\rm e0}$ (equal to the
cross-correlation of \mgii systems and CMASS galaxies), times $r_{\rm p}(1+z)$ for the mean subtraction method, the variable smoothing method and measurements from \protect\cite{Zhu2013} respectively.
The thick solid black line is the MultiDark model prediction described in
\protect\cite{Nuza2013}. The solid blue, the red dashed and the green dashed-dotted lines are the fit to $W_{\rm e}$ for the mean subtraction method, the variable smoothing method and measurements from \protect\cite{Zhu2013} respectively. The ratio of the
each of these lines with the thick solid black line is the ratio of bias factors,
 $b_{\mathrm{Mg}}/b_{\mathrm{g}}$.}
	\label{fig: correlation function comparison}
\end{figure*}

  We now use equation \ref{eq: delta MgII} to infer the bias
factor of the \mgii systems, as
\begin{equation}
 \label{eq: delta MgII 2}
 b_{\mathrm{Mg}}= b_{\rm g}\, \frac{c\, W_{\rm e}(r_{\rm p})(1+z)}
 {\tau_{\rm e0}\lambda_{\rm \mgii}\, w_{\rm gg}(r_{\rm p})\,\, H(z)} ~.
\end{equation}
Note that the factor $(1+z)/H(z)$ appears because of our convention that
the effective equivalent width $W_{\rm e}$ is measured in \angs, whereas
$w_{\rm gg}$ is assumed to have been transformed to comoving length units.
Instead of fitting our \mgii\!\!-galaxy cross-correlation measurement with
a power-law dependence with impact parameter (as in equation
\ref{eq: absorption model}), we can directly fit the functional form
that is determined from the observed projected galaxy correlation
function, assuming that the shape is the same. To do this,
we use the projected galaxy autocorrelation function obtained in \cite{Nuza2013}
from the BOSS DR9 catalog of CMASS galaxies, and their prediction for the
galaxy correlation function based on assigning galaxies to halos and subhalos in
their MultiDark simulation. The measurements of \cite{Nuza2013}
are represented as black triangles with errorbars in figure
\ref{fig: correlation function comparison}, and their model is shown
as the thick black line (given in their figure 6 and table B1; note that
we have corrected for the different cosmological model
they use, with a present matter density $\Omega_m=0.27$ instead of our
value $\Omega_{\rm m}=0.3$). Blue triangles, red squares and green circles are the mean equivalent width
$W_{\rm e}(r_{\rm p})$ times the factor $(1+z)/H(z)/\tau_{\rm e0}$ (equal to the
cross-correlation of \mgii systems and CMASS galaxies), times $r_{\rm p}(1+z)$ for the mean subtraction method, the variable smoothing method and measurements from \cite{Zhu2013} respectively.

  The galaxy bias factor in the model of \cite{Nuza2013} shown as the
dashed black line is $b_{\rm g}=2.00\pm0.07$. Note that this value is
lower than that obtained by \cite{Guo2013}, $b_{\rm g}=2.16\pm0.01$, for
the average CMASS galaxy. Using the value given by \cite{Guo2013} would
lead to a larger measured value of the bias factor of the \mgii systems. 
As explained in \cite{Guo2013}, the value of the galaxy bias increases
with luminosity and redshift, and it can also depend 
on the range of scales used to fit its value. Here we use the galaxy bias
value and the projected galaxy autocorrelation function of \cite{Nuza2013},
but the discrepancy with the higher value obtained by \cite{Guo2013}
needs to be resolved to remove this source of uncertainty on the
derived bias value of the \mgii absorption systems.
 
  We now redo the fit to the measured $\delta\tau_{\rm e}(r_{\rm p},v)$ profiles
presented in section \ref{sec: Results}, after replacing equation
(\ref{eq: absorption model}) by
\begin{displaymath}
	\delta{\tau}_{\rm e}(r_{\rm p},v) =
 \frac{b_{\mathrm{Mg}}\, \tau_{\rm e0}}{b_g \sqrt{2\pi}(1+q)}\, 
 \frac{ w_{\rm gg}(r_{\rm p}) H(z)}{(1+z)\sqrt{\sigma_0^2 + (xHr_{\rm p})^2}} \,\times
\end{displaymath}
\begin{equation}
	\label{eq: wgg absorption model}
 \left[{\mathrm e}^{-\left(v-v_{1}\right)^{2}/2\sigma^{2}} + q\,
  {\mathrm e}^{-\left(v-v_{2}\right)^{2}/2\sigma^{2}}\right] ~.
\end{equation}
That is to say, we replace the power-law dependence of $W_{\rm e}$ on impact
parameter by the model to the observed $w_{gg}$ of \cite{Nuza2013}. All
the parameters except for $b_{\mathrm{Mg}}$ are kept fixed to the best fit obtained
in section \ref{sec: Results}, and a new fit is obtained by matching only
$b_{\mathrm{Mg}}$. The function $w_{\rm gg}(r_{\rm p})$ is computed at our impact parameter
bins by linear interpolation of the values of the model of \cite{Nuza2013}.
For convenience the fit is performed in the equivalent width space, namely,
\begin{equation}
	\label{eq: wgg absorption model 2}
	W_{e}\left(r_{p}\right) = \tau_{e0} \frac{\lambda_{\mgii}}{c}\frac{b_{\rm Mg}}{b_{g}}\frac{w_{gg}\left(r_{p}\right)H(z)}{1+z}
\end{equation}
The results we obtain for the \mgii absorption bias factor are
\begin{equation}
 b_{\mathrm{Mg}\,\mathrm{mean\,subtraction}}=2.33\pm 0.19 ~,
\end{equation}
\begin{equation}
 b_{\mathrm{Mg}\,\mathrm{variable\,smoothing}}=1.14\pm 0.36 ~,
\end{equation}
and the resulting fit for $W_{\rm e}(r_{\rm p}) r_{\rm p}$ are shown as
the solid blue line for the mean subtraction method and the dashed red line for the variable smoothing method in figure \ref{fig: correlation function comparison}
(these lines are simply the rebinned thick solid black line shifted by the factor
$b_{\mathrm{Mg}}/b_{\rm g}$). The error on this bias factor includes
only the uncertainty of the fit that assumes the bootstrap errors in our
stacked spectra, and does not include the error of $\tau_{\rm e0}$,
assumed to be $\tau_{\rm e0}=5.0\times 10^{-4}$.
For completeness we also repeat the fit using \cite{Zhu2013} datapoints. The value we obtain is
\begin{equation}
 b_{\mathrm{Mg}\,\mathrm{ZHU13}}=0.78\pm 0.05 ~,
\end{equation}
Note that the error here is computed not from the bootstrap errors but directrly from the $\chi^{2}$ fitting instead. We neglect the fact that the datapoints are not independent.

  We find a huge difference for $b_{\mathrm{Mg}}$ for the different methods. Thus, we stress once again the importance of the quasar continuum estimate. Getting the estimate right is crucial for the measurement of the bias and failing to do so may lead to really different results.

  Our measurement for the mean subtraction method is discrepant from the previously reported values by
\cite{Gauthier2009} of $b_{\mathrm{Mg}}=1.36\pm 0.38$, and by
\cite{Lundgren2009} of $b_{\mathrm{Mg}}=1.10\pm 0.24$. 
Our result is closer to the bias factor measured for the galaxies,
implying that most of the \mgii systems are associated to massive
galaxies like the CMASS ones or even more massive.
On the other hand, the measurement for the variable smoothing method is compatible with the previous ones. Finally the measurement using \cite{Zhu2013} datapoints is also not compatible with the reported values but for the opposite reason. It is too low.

However, as we discussed in section \ref{sec: Results} and in the Appendix \ref{sec: Test}, both the variable smoothing method and the method used in \cite{Zhu2013} underestimate the observed mean equivalent width. We have presented proof that it is indeed true for the variable smoothing case. We also argued that this is also likely to be true for the method used in \cite{Zhu2013}. 
This means that the 'correct' value should be the $2.33\pm 0.19$ obtained for the mean subtraction method even though it is not compatible with the values reported in \cite{Gauthier2009} and \cite{Lundgren2009}.

Note that our measurement of the bias factor remains
subject to systematic uncertainties that will need to be improved:
the determination of $\tau_{e0}$, the use of a wider velocity interval
for determining the quasar continuum and the mean \mgii absorption
compared to the ones used in this paper, and the use of a better
modeling of the cross-correlation that includes redshift distortions
in the regime of large impact parameters, and a more general density
profile of \mgii absorbers in halos of different mass.

  One possible reason for our high value of the \mgii bias is the degeneracy between $\tau_{\rm e0}$ and the \mgii bias. What we are actually fitting is the product of both. We can only recover the \mgii bias once we fix the value of $\tau_{\rm e0}$. This means that an underestimation of $\tau_{\rm e0}$ will result in an overestimation of the \mgii bias. Thus, a more robust measurement of $\tau_{\rm e0}$ is required.
Another possible reason for our high value of the \mgii bias is that the
bias may decrease with the equivalent width of the \mgii absorbers,
as found both by \cite{Lundgren2009} and \cite{Gauthier2009}. Our
method includes all the \mgii absorbers and measures their average
bias, weighting each absorber by its equivalent width. This average
bias would be larger than the one for strong, individually detected
\mgii lines if the weak absorbers are associated with massive halo
environments, whereas absorbers of high equivalent width occur in
galaxies hosted by low mass halos. Yet another possible explanation is
our use of a limited velocity range for evaluating the projected
cross-correlation of \mgii absorption and galaxies. We note that the
high value of the bias we obtain is driven by the last point (at
largest impact parameter) in figure \ref{fig: correlation function comparison}.
This point might be too high because linear redshift distortions have
increased the density of \mgii absorbers in the interval used for
integration, and decreased them in the interval used for continuum
fitting. The projected cross-correlation should not be affected by these
redshift distortions when it is computed by integrating over the whole
line of sight, but at the largest impact parameters our integrating
intervals are probably not large enough. This systematic effect can only
be addressed by improving the continuum fitting method and the model of
the cross-correlation in future work.

  We now compare our results for the \mgii\!\!-galaxy
cross-correlation with those of \cite{Zhu2013}. The modeling of 
\cite{Zhu2013} involves a free parameter, which they designate as
$f_{\mathrm{\mgii}}$, that reflects the
gas-to-mass ratio in halos (ignoring the degree of saturation of the
absorption lines), and they assume a fixed density profile for the
absorbers in halos. \cite{Zhu2013} do not relate this parameter to the
mean absorption of the field population of \mgii absorbers, so
they fit the amplitude of the cross-correlation with this unconstrained,
free parameter. They determine a characteristic host halo mass for the
\mgii absorbers of $M_{\rm h}\simeq 10^{13.5} \msun$ based on the presence of a
feature in the shape of the cross-correlation at $r_{\rm p}\sim 1$ Mpc that
reflects a transition from the 1-halo term to the 2-halo term in their
modeling. The weakness of this feature implies a poor determination of
this characteristic halo mass (see the contours in their figure 6,
showing the large degeneracy with the $f_{\mathrm{\mgii}}$ parameter).
Their determination of this halo mass is therefore highly dependent on
their model of the halo density profile, and does not generally match
the total observed abundance of \mgii absorbers.
Moreover, the \mgii absorbers are
likely hosted in halos with a very broad mass range, which should cause
a smoothing of any feature due to the transition from the 1-halo to the
2-halo term. The specific density profile of absorbers they assume has
not been tested and is not theoretically well motivated.
\mgii absorbers can be distributed in halos differently from
galaxies, depending on the physics of gas cooling and galactic
winds in halos.
Auto-correlations and cross-correlations of galaxies of different types
have been found to have widely different shapes \citep[see, e.g., figures
15 and 16 in][]{Zehavi2011} which do not always possess the clear
feature that is predicted for a tracer that follows the dark matter
profile in halos of a specific mass. Therefore, we think there is
insufficient evidence for the presence of any feature in the \mgii\!\!-galaxy
cross-correlation that may be used to determine a characteristic host
halo mass.

  Instead, we propose that the {\it amplitude} of this
cross-correlation should be related to the field population of \mgii
absorbers (e.g., through the effective optical depth $\tau_{\rm e0}$
that we have introduced), and can then be used to determine the mean
bias factor $b_{\mathrm{\mgii}}$, which should be equal to the
mean bias factor of the host halos of the absorbers (weighted by their
rest-frame equivalent width), and is robustly defined even if the range
of host halo masses is very broad.

\subsection{The ratio of \mgii-absorbing gas to the total mass}
 \label{subs: gasmass ratio}

  Measurements of the average \mgii absorption around galaxies can be
compared with mass measurements averaged in the same way obtained from
weak gravitational lensing. This comparison was done in \cite{Zhu2013}
to obtain an estimate for the ratio of gas mass to total mass in the
halos around the CMASS galaxies of the BOSS survey. We now examine this
question to point out a number of uncertainties in this derivation.

  In general, the total column density of \mgii in an individual
absorber, $N_{\mathrm{\mgii}}$, is related to its integrated optical depth
according to
\begin{equation}
 N_{\mathrm{\mgii}}={m_{\rm e} c^2\over \pi {\mathrm e}^2}\,
{W_\tau\over f \lambda_{\mathrm{\mgii}}^2} = 1.13\times 10^{20} \,
{W_\tau \angs \over f \lambda_{\mathrm{\mgii}}^2}\, \cm^{-2} ~.
\end{equation}
where $e$ and $m_{\rm e}$ are the electric charge and mass of the electron,
and $f=0.921$ is the total oscillator strength of the \mgii doublet.
The integrated optical depth is
\begin{equation}
 W_\tau = \int {\rm d}\lambda \tau(\lambda) ~,
\end{equation}
where the integration is performed over a wavelength range that includes
the entire absorption profile. However, the only quantity
that is observed is the equivalent width,
\begin{equation}
 W = \int {\rm d}\lambda \left[ 1-e^{-\tau(\lambda)} \right] ~.
\end{equation}
When the optical depth of the absorber is much less than unity over
the whole wavelength range, the absorber is unsaturated and
$W_\tau \simeq W$. Otherwise, the column density is not directly
measurable simply from the equivalent width. We now define an average
saturation level for the population of absorbers, $\bar S$, as
\begin{equation}
 \bar S =  { \int {\rm d}W \, 
  (\partial^{2} \mathcal{N}/\partial W\partial z) \, W_\tau  \over
  \int {\rm d}W (\partial^{2} \mathcal{N}/\partial W\partial z) \, W } ~.
\end{equation}
Defining also $x_{\mathrm{\mgii}}$ as the fraction of magnesium atoms in the
absorbing gas that are in the \mgii ionized species, $g_{\mathrm{Mg}}$ as the
fraction of magnesium in the absorbers that is in the gas phase (as
opposed to dust grains), and $Z_{\mathrm{Mg}}$ as the magnesium mass fraction
compared to that of the Sun (we use a solar magnesium abundance by
mass of $7.0\times 10^{-4}$, and a magnesium mass
$m_{\mathrm{Mg}}=4.07\times 10^{-23} \, {\rm g}$), we obtain that the total gas
mass surface density in the \mgii absorbers is
\begin{equation}
\label{eq: gsmd}
 \Sigma_g (r_p)= 9.15\times 10^{-7}
{\bar S\over x_{\mathrm{\mgii}}\, g_{\mathrm{Mg}}\, Z_{\mathrm{Mg}} } \,
{W_{\rm e}(r_{\rm p})\over \text{\angs} } {\,\rm g} \cm^{-2} ~.
\end{equation}

  The total mass surface density around a CMASS galaxy in the BOSS
sample has also been measured using weak gravitational lensing. We can
therefore obtain the ratio of gas in \mgii absorbers to the total mass
by combining the two observational measurements. We use the recent
measurement by \cite{Miyatake2013} based on weak
lensing measurements in the CFHTLenS survey area. As an example, we
compute the gas-to-mass ratio at a comoving projected radius of
$r_{\rm p}(1+z)=3$ Mpc. The differential surface density measured by
\cite{Miyatake2013} at this radius is $\Delta\Sigma = \bar\Sigma -
\Sigma \simeq 2 \msun\, {\rm pc}^{-2}$ (see their figure 7). Near this
radius, $\Delta\Sigma(r_{\rm p})$ is falling with radius roughly as
$r_{\rm p}^{-1}$, so the mean surface density within $r_{\rm p}$
is $\bar{\Sigma} \simeq 2 \Sigma$, and we can therefore use $\Sigma
\simeq \Delta\Sigma$. At this same radius, $W_{\rm e}\simeq 0.03 \angs$,
and substituting this value in equation (\ref{eq: gsmd}), it produces
$\Sigma_{\rm g}/\Sigma \simeq 8\times 10^{-5} \bar S/(x_{\mathrm{\mgii}}\,
g_{\mathrm{Mg}}\, Z_{\mathrm{Mg}})$. Using the mean
ratio of baryons to total matter in the universe of
$\Omega_{\rm b}/\Omega_{\rm m} = 0.17$, the fraction of baryons in the \mgii
absorbing gas would be
$\Sigma_{\rm g}/\Sigma_{\rm b} \simeq 5\times 10^{-4} \bar S/(x_{\mathrm{\mgii}}\,
g_{\mathrm{Mg}}\, Z_{\mathrm{Mg}})$.

  Therefore, the fraction of baryons in the \mgii clouds can be a small
one even if the mean metallicity is relatively low. However, the mean
saturation parameter $\bar S$ is likely much larger than unity, so it is
possible that the \mgii absorbers account for an important fraction of
the baryons in galactic halos, and for the accreting material that fuels
the star formation rate. We note that any further comparison of the
detailed radial profiles of $W_e(r_p)$ and $w_{gg}(r_p)$ cannot be
reliably used to infer a profile of the gas-to-mass ratio, because
$Z_{\mathrm{Mg}}$ is likely to vary with $r_p$, since the heavy elements
must have originated from galactic winds, and the values of $\bar S$,
$x_{\mathrm{\mgii}}$ and $g_{\mathrm{Mg}}$ may also vary substantially
with $r_p$. The mean value of the gas-to-mass ratio is still highly
uncertain because of the unknown value of
$\bar S/(x_{\mathrm{\mgii}}\, g_{\mathrm{Mg}}\, Z_{\mathrm{Mg}})$.

\section[Conclusions]{Summary and conclusions} \label{sec: Conclusions}
In this paper we have used the \mgii line to measure the cross-correlation 
of \mgii absorption and galaxies in BOSS. The large size of the samples we 
use (SDSS DR7 quasar catalog as background sample and DR11 CMASS galaxy 
catalog as foreground sample) enables a statistical approach to detect \mgii 
absorption that is too weak to be detected individually and would otherwise 
be missed. We present a method to estimate the quasar continuum 
designed for this type of measurements and compare our results with those 
obtained by a more typical continuum estimate. Our main results can be 
sumarized as follows:

\begin{itemize}
 \item The method to fit the quasar continuum is crucial to measure the
mean \mgii equivalent width as a function of impact parameter. Methods
that use the observed flux in the spectral region near the \mgii
line wavelength at the galaxy redshift to determine the continuum
suffer from a systematic bias, because the absorption from individually
undetected systems inevitably lowers the continuum estimate and causes
an underestimate of the mean absorption. The tests presented in the Appendix
\ref{sec: Test} show that our mean subtraction method does not suffer
from any systematic effect to the extent that we are able to discern.

 \item We find that the cross-correlation of \mgii absorption and CMASS
galaxies follows the shape of the CMASS galaxies auto-correlation at
large scales. We use the CMASS auto-correlation model from
\cite{Nuza2013} and the measured galaxy bias factor to derive a bias
factor of \mgii absorbers of $b_{\mathrm{Mg}}=2.33\pm 0.19$. This value
is substantially larger than the previous measurements by
\cite{Gauthier2009} and  \cite{Lundgren2009}. This discrepancy may be
due to a real difference, because our measurement includes a
contribution from weak \mgii absorption systems which may be more
strongly clustered than strong absorbers, and may also be affected by
our imperfect determination of the projected cross-correlation at large
impact parameters owing to our limited integrating range. More accurate
measurements and better modeling will be necessary to clarify this question.

\end{itemize}

\vspace{6mm}

\section*{Acknowledgments}
Funding for SDSS-III has been provided by the Alfred P. Sloan
Foundation, the Participating Institutions,
the National Science Foundation, and the U.S. Department of Energy
Office of Science. The SDSS-III web site is http://www.sdss3.org/.
IP and JM have been supported in part by Spanish grants
AYA2009-09745 and AYA2012-33938.

SDSS-III is managed by the Astrophysical Research Consortium for the Participating
Institutions of the SDSS-III Collaboration including the University of Arizona, the Brazilian Participation Group, Brookhaven National Laboratory, University of Cambridge, Carnegie
Mellon University, University of Florida, the French Participation Group, the German Participation
Group, Harvard University, the Instituto de Astrofisica de Canarias, the Michigan
State/Notre Dame/JINA Participation Group, Johns Hopkins University, Lawrence Berkeley
National Laboratory, Max Planck Institute for Astrophysics, Max Planck Institute for
Extraterrestrial Physics, New Mexico State University, New York University, Ohio State University,
Pennsylvania State University, University of Portsmouth, Princeton University, the
Spanish Participation Group, University of Tokyo, University of Utah, Vanderbilt University,
University of Virginia, University of Washington, and Yale University.

\bibliographystyle{apj}
\bibliography{MgII_project}{}

\appendix
\section{Tests of the continuum fitting methods}  \label{sec: Test}

  The method to fit the quasar continuum is a crucial part of the
measurement of the mean \mgii absorption equivalent width as a function
of impact parameter from a galaxy, $W_\tau(b)$, presented in
this paper. The two methods we have used produce a different
result, which is also different from the result reported by
\cite{Zhu2013}. It is therefore important to perform tests on
these methods that can reveal the presence of systematic errors in the
$W_\tau(b)$ estimates. This appendix presents the results of three
tests. The first one (section \ref{subs: systematics}) checks for any
systematic mean absorption that might be artificially introduced by the
continuum fitting method when there is no correlation between \mgii
absorbers and galaxies. The second one
(section \ref{subs: detected systems}) verifies that the correct
equivalent width of an individually detected \mgii absorption system in a spectrum
is correctly recovered. Finally, section \ref{subs: undetected systems} 
reveals the effect on the
fitted continuum of the presence of weak absorbers that are individually
undetected, and the way these absorbers can bias the estimate of the mean
equivalent width.

\subsection{Systematic errors in the absence of correlations}
 \label{subs: systematics}

  One might suspect that a small average absorption (either positive or
negative) is artificially introduced by the method to fit the continuum,
even when the regions selected to search for absorbers are completely
random and should therefore have no average absorption. This might
happen if the quasar continuum is systematically overestimated or
underestimated, depending in a complex manner on the varying noise
properties and the shape of the true quasar continuum. To test for this
possibility, we have remeasured the mean \mgii equivalent widths after
rotating the right ascension coordinate of all the quasar by 5 and 10
degrees, in the two possible directions, and after increasing and decreasing
the redshift of the galaxies by 0.05.
These separations are large enough 
to make any residual cross-correlation of galaxies
and \mgii absorbers completely negligible, so the measured correlation
should be consistent with zero. Note that this procedure ensures that
the auto-correlations that are present among the \mgii absorbers,
quasars and galaxies are preserved, so their contribution to the
measurement errors of the cross-correlation is the same.

\begin{figure*}
	\includegraphics[width=\textwidth]{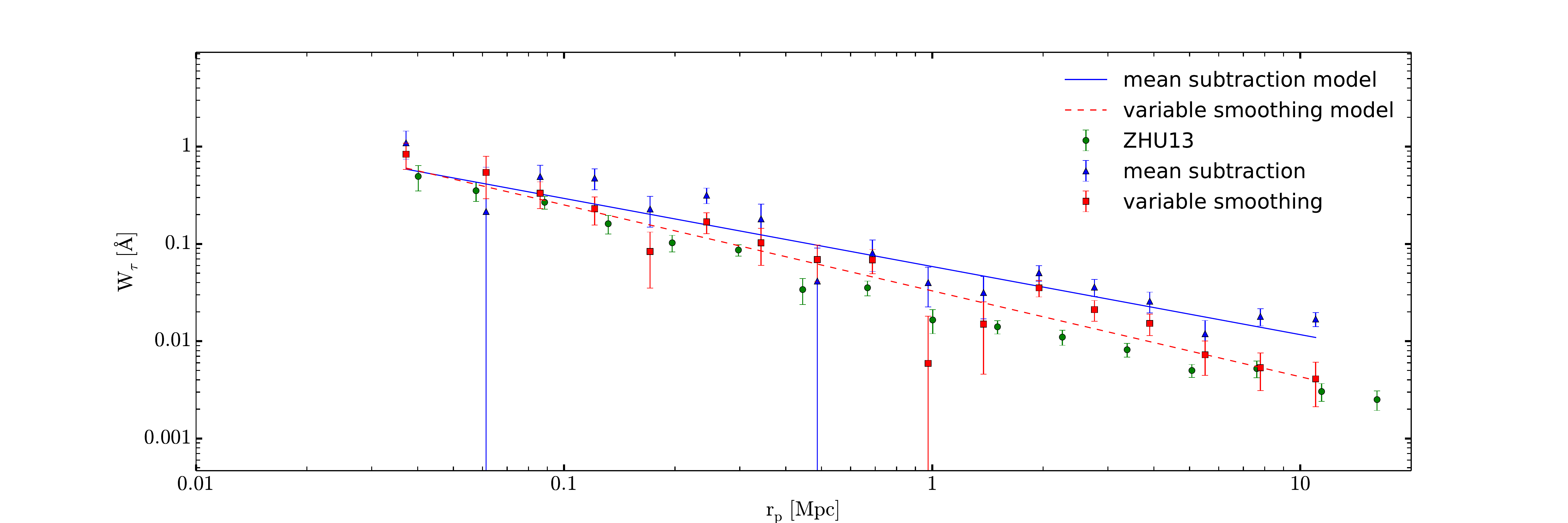}
	\caption{Ratio of the mean equivalent width in stacked spectra
to the best-fit power-law model prediction, for the real data sample
(big red circles), and for the average of 6 mock samples (blue stars). 
Errorbars are computed with the bootstrap
method. This ratio should be consistent with zero for the mock sample
in the absence of systematic errors, and with unity for the real data
if the power-law model provides a good fit to the data. }
	\label{fig: test-a}
\end{figure*}

  The result of this test is shown in figure \ref{fig: test-a}, in the
left panel for the mean subtraction method and the right panel for the
variable smoothing method. The
real data set is shown as big red circles with errorbars, after dividing by
the best-fit power-law model that is described in Section
\ref{sec: absorption model} and plotted in figure \ref{fig: ew_plot}.
The average of 6 mock data sets are shown also after dividyng by the same model as 
blue stars.
In the absence of any systematics, the
mean absorption in the mock data sets is expected to be zero, whereas
the real data should produce a ratio to the best-fit model that is
consistent with one. The results do not show any systematic errors for
the mean subtraction method. There appears to be a small negative systematic
absorption that is introduced by the variable smoothing method, as indicated
by the negative values of the mock sample in the right panel, at large
impact parameters (where the mean equivalent width can be measured
with the smallest errorbar). This average negative absorption is approximately
equal to the best-fit model prediction at $\sim 10$ Mpc, or $\sim 0.3
\kms$ (see figure \ref{fig: ew_plot}). This result may be due to some subtle effect in
the variable smoothing method that introduces a small bias by systematically
underestimating the continuum in the absence of real absorption lines,
possibly due to occasional false identifications of noise spikes as
real absorbers. As we shall see below, the variable smoothing
method is actually affected by a more serious systematic error that
partially eliminates the contribution of weak absorption systems to the
mean equivalent width.

\subsection{Tests of the equivalent width measurement for individually
 detected systems}\label{subs: detected systems}

\begin{figure*}
	\includegraphics[width=\textwidth]{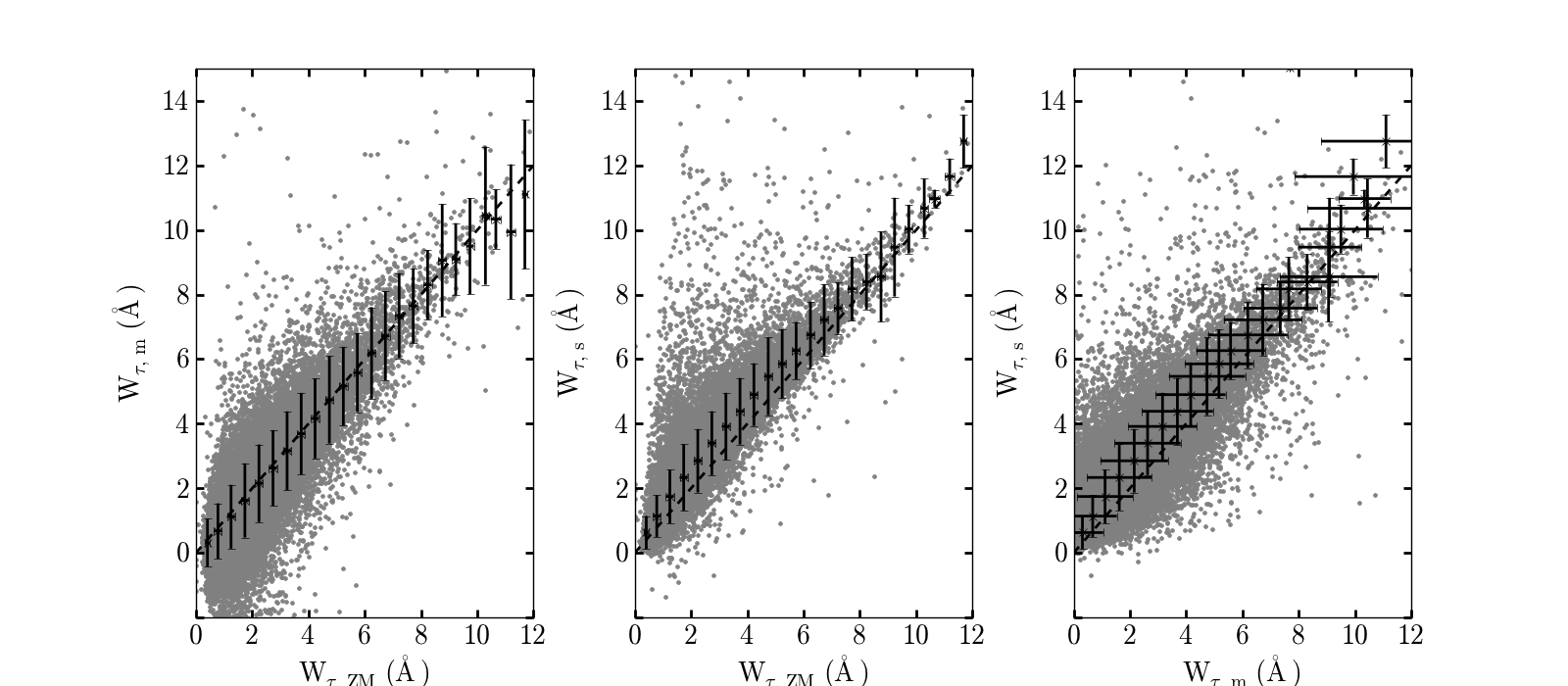}
	\caption{Comparison of the equivalent width estimates from
different methods (see Section \ref{sec: Stacking procedure}). From left
to right, equivalent widths measured using the mean
subtraction method, $W_{\rm m}$, versus the value of the Zhu \& Menard
catalogue, $W_{\rm ZM}$; variable smoothing method value, $W_{\rm s}$, versus
$W_{\rm ZM}$; and $W_{\rm s}$ versus $W_{\rm m}$. Black points are the average
values in bins of $\Delta W=0.5 \angs$ in the horizontal axis, with
the dispersion in each bin indicated by the errorbars. One-to-one
correspondence is marked by the black dashed line for visual guidance.}
	\label{fig: ew scatter}
\end{figure*}

  We now test that the mean equivalent width measured for absorbing
systems that are individually clearly detected above the noise agrees
with other well established methods. For this purpose, we use the \mgii
absorber catalogue of \cite{Zhu&Menard2013b}, which contains $35,752$
absorption systems from the SDSS DR7 quasar spectra sample. The
integrated equivalent widths we obtain for systems in this catalogue,
using our two methods of mean subtraction and variable smoothing, are
compared with the equivalent widths provided in the catalogue in figure
\ref{fig: ew scatter}. There is a large scatter in the equivalent widths 
obtained with different methods. This result
is not surprising, because the noise can change the determination
of the continuum in random ways in different methods. In particular, in
the method of the mean subtraction, the equivalent width is obtained by
integrating the absorbed fraction over a wide interval around the
absorber, according to equation (\ref{eq: EW direct}), adding noise to
the estimate. However, the average of the equivalent width estimator in
our mean subtraction method, shown by the black points in figure
\ref{fig: ew scatter}
(with an rms dispersion indicated by the errorbars), agrees very well
with the equivalent width provided by the \cite{Zhu&Menard2013b} catalogue. The
variable smoothing method apparently suffers from a bias causing a $10$ to
$20\%$ increase of the average of the equivalent width (see black points
in middle panel of figure \ref{fig: ew scatter}), which may be due to a
tendency of this method to overestimate the continuum level around
detected absorption lines.

\subsection{Impact of individually undetected systems on the mean
equivalent width}\label{subs: undetected systems}

  We now test how the presence of weak absorption systems that cannot be
individually detected in a single spectrum but contribute to the mean
equivalent width as a function of impact parameter from a galaxy can
bias the estimate of the quasar continuum in the different methods we
use. For this purpose, artificial absorbers are introduced in a
spectrum, and then we refit the continuum and measure the change in the
measured equivalent width.

  As an illustrative example, we have selected a set of 10 random quasar
spectra, and we have introduced absorbers at 10 random redshift values
and computed the average values for the recovered width.
The absorbers are inserted with a double Gaussian profile in the optical depth, as expected for the \mgii doublet,
\begin{displaymath}
  \tau(v) = \frac{W_{0}}{\sqrt{2\pi}\sigma} 
\left( {2\over 3} \exp\left[-\left(v-v_{1}\right)^{2}/2\sigma^{2}\right] + \right.
\end{displaymath}
\begin{equation}
	\label{eq:itau}
\left. + {1\over 3} \exp\left[-\left(v-v_{2}\right)^{2}/2\sigma^{2}\right]\right) ~,
\end{equation}
where $W_0$ is the total equivalent width of the doublet and $\sigma$ is
velocity dispersion. The zero velocity is conventionally chosen to be
the central position of the \mgii line for an unsaturated line, so that
$v_{1}=-256.05 \kms$ and $v_{2}=513.28 \kms$. In the absence of any
inserted absorber, the continuum determined in this spectrum is $c(v)$,
the flux is $f(v)$, and the transmitted fraction is $F(v)= f(v)/c(v)$.
This results in a certain value of the integrated equivalent width $W$
measured over the interval $-2000 \kms < v < 2000 \kms$, by integrating
$F(v)$ over this range. To insert the absorber, the spectral flux is
modified according to 
\begin{equation}
	f'(v)=f(v)-c(v)\left(1-\exp\left[-\tau(v)\right]\right)~.
\end{equation}
Then, a new continuum $c'(v)$ is determined with the new flux, and a
new transmitted fraction $F'(v)=f'(v)/c'(v)$ is derived. Finally, the
new equivalent width $W'$ is determined by integrating $F'$ over the
same velocity interval.

\begin{figure*}
	\includegraphics[width=\textwidth]{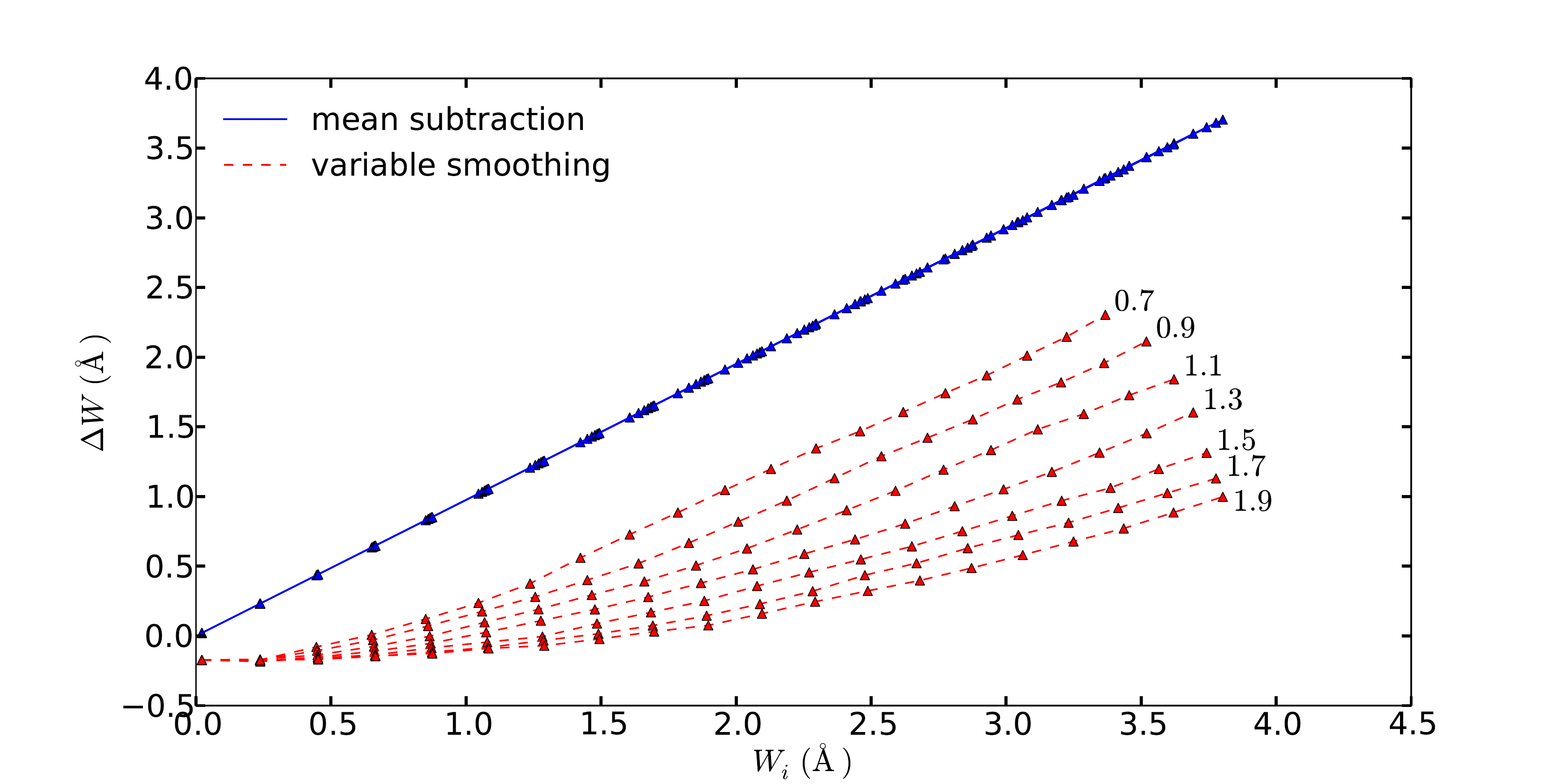}
	\caption{Change in the measured equivalent width, $\Delta W$,
caused by the insertion of an absorbing system with equivalent width
$W_i$, for the mean subtraction method (solid blue lines) and the variable
smoothing method (dashed red lines). Results are shown as a function of the
velocity dispersion $\sigma$, with values indicated to the right of
the lines in \angs. The blue lines all nearly coincide at $\Delta W = W_i$.
Points show the values of $W_i$ for which
$\Delta W$ has been computed; the sudden changes in the red lines
indicate discontinuities in the variable smoothing method as the inserted
line becomes detected or covers different pixels, which
causes a change in the continuum estimate.}
	\label{fig: introduced absorption}
\end{figure*}

  The change in equivalent width caused by the insertion of an
absorber, $\Delta W = W'-W$, is plotted in figure
\ref{fig: introduced absorption} as a function of the equivalent width
$W_i$ of the inserted absorber, obtained by integrating
$1-\mathrm{exp}[-\tau(v)]$ over the same velocity interval that is used for
determining $W$ and $W'$ ($W_i$ is nearly equal to $W_0$ in equation
\ref{eq:itau}, except that the integrating interval does not extend to
infinity). The different lines correspond to
different values of the absorber velocity dispersion, $\sigma$. The solid
blue lines represent the mean subtraction method, and they coincide
precisely with $\Delta W = W_i$ for all values of $\sigma$. The result,
as expected, is that the continuum determined by this method is
unaffected by the presence of the absorbers that
have narrow widths compared to the chosen integrating interval width of
$4000 \kms$. The reason is that, in the mean subtraction method, the
continuum $c'(v)$ is determined using only the measured flux outside of
this interval.

  On the other hand, the variable smoothing method (dashed red lines in figure
\ref{fig: introduced absorption}) is strongly biased to lowering the
estimated continuum in response to the presence of a weak absorption
system. The result is that the change caused by the absorber in the
estimated equivalent width, $\Delta W$, can be much less than the true
value, and the difference is a complex function of the equivalent width
$W_0$, the velocity dispersion and the signal-to-noise ratio of the spectrum.
The underestimate of the continuum level is naturally smaller for
narrower lines (lower $\sigma$), because the lines are detected and
eliminated from the continuum estimation for lower values of $W_0$.

  To summarize, the three tests of our continuum fitting methods
presented in this Appendix demonstrate that the variable smoothing method suffers
from several systematic errors. The first test shows that a small
negative absorption, of equivalent width $\sim -0.3\kms$, is induced
where there is none. The second test indicates that the equivalent width
of strong, detected systems is overestimated by $10-20\%$. Finally, the
third test shows that for weak systems, the continuum is systematically
underestimated, thereby strongly reducing the contribution of these
systems to the measured equivalent width. However, the mean subtraction
method successfully passes all these tests and should therefore provide
a reliable estimate of the stacked equivalent width as a function of
impact parameter.

\end{document}